\documentclass{aa}  
\usepackage{comment}
\usepackage{csquotes}
\usepackage{IEEEtrantools}
\usepackage{longtable}
\usepackage{graphicx}
\usepackage{pdflscape}
\usepackage{caption}
\usepackage{color}
\usepackage{multirow}
\usepackage{cancel}
\usepackage{txfonts}
\def\arcdeg{^\circ}
\def\degr{^\circ}
\def\pr{P$_\text{R}$~}
\def\tr{$\theta_\text{R}$~}
\def\pp{P$_\text{P}$~}
\def\tp{$\theta_\text{P}$~}
\def\ph{P$_\text{H}$~}
\def\th{$\theta_\text{H}$~}
\def\ticm{$\theta_\text{ICM}$~}
\def\bpos{B$_{\mathrm{POS}}$~}
\def\arcmin{^\prime}
\def\arcsec{^{\prime\prime}}

\begin{document}

   \title{Tracing the magnetic field morphology of the LDN 1172/1174 cloud complex}

   \subtitle{}
   \titlerunning{Magnetic field study toward the LDN 1172/1174}
   \authorrunning{Piyali Saha et. al}

   \author{Piyali Saha\inst{1,2,3}
	\and
	Maheswar G.\inst{1}
	\and
	Ekta Sharma\inst{1,4}
	\and
	Chang Won Lee\inst{5,6}
	\and
	Tuhin Ghosh\inst{7}
	\and
	Kim, Shinyoung \inst{5}
}

\institute{Indian Institute of Astrophysics (IIA), Sarjapur Road, Koramangala, Bangalore 560034, India\\
	\email{s.piyali16@gmail.com}
	\and
	Satyendra Nath Bose National Centre for Basic Sciences (SNBNCBS), Salt Lake, Kolkata-700 106, India
	\and
	Pt. Ravishankar Shukla University, Amanaka G.E. Road, Raipur, Chhatisgarh 492010, India
	\and
	Department of Physics and Astrophysics, University of Delhi, Delhi 110007, India
	\and
	Korea Astronomy, and Space Science Institute (KASI), 776 Daedeokdae-ro, Yuseong-gu, Daejeon 305-348, Republic of Korea
	\and
	University of Science and Technology, Korea, 217 Gajeong-ro, Yuseong-gu, Daejeon 34113, Republic of Korea
	\and
	School of Physical Sciences, National Institute of Science Education and Research, HBNI, Jatni 752050, Odisha, India}

\date{Received ....; accepted ....}

 
  \abstract
   {The LDN 1172/1174 cloud complex in the Cepheus Flare region presents a hub-filament structure with the reflection nebula, NGC 7023, illuminated by a Herbig Be star, HD 200775, which consists of the hub with a $\sim$5 pc long narrow filament attached to it. Formation of a sparse cluster of low- and intermediate-mass stars is presently taking place in the hub.}
   {The aim of this work is to map the magnetic field geometry of LDN 1172/1174 to understand the role played by the field lines in the formation of the molecular cloud.}
   {Unpolarized background stars can be used to measure dichroically polarized light produced by the magnetically aligned grains present in molecular clouds. As these dust grains get aligned with their minor axes parallel to the ambient magnetic field, the polarization measurements can provide the plane-of-sky component of the magnetic field. We made R-band polarization measurements of 249 stars projected on the entire LDN 1172/1174 cloud complex to map the geometry of the magnetic field of this region.}
   {The magnetic field geometry constructed from our R-band polarization measurements is found to be parallel to the elongated structure inferred from the column density distribution of the cloud produced using the \textit{Herschel} images. Our R-band polarization measurements are found to be in good agreement with those obtained from \textit{Planck}. There is evidence of a possible distortion of the magnetic fields toward the northwestern part of the cloud by HD 200775. The magnetic field strength is estimated as $\sim$30 $\mu$G. The estimated star formation rate (SFR)/mass of 2.0$\pm$1.3 \%Myr$^{-1}$ and 0.4$\pm$0.3 \%Myr$^{-1}$ for LDN 1172/1174 and the neighboring cloud complex, LDN 1147/1158, respectively, are found to be consistent with the mean SFR/mass found for the clouds with magnetic field orientations parallel and perpendicular to their elongated structures, respectively. These results support earlier findings that the clouds with magnetic field lines parallel to their long axes seem to have higher SFRs compared to those with the magnetic field orientation perpendicular to the cloud elongation.}
   {}


   \maketitle
%

\section{Introduction} \label{sec:intro}

Star formation is often found to be associated with the densest regions of molecular clouds that are elongated and filamentary in shape. In fact, results obtained with the \textit{Herschel} space observatory \citep{2010A&A...518L...1P} of nearby interstellar clouds, as well as in more distant and massive clouds, reveal the ubiquity of complex networks of filaments \citep{2010A&A...518L.102A, 2010A&A...518L.100M, 2011A&A...533A..94H}. How molecular clouds form and then evolve into filaments and cores and finally collapse to form stars is still a puzzle. Turbulence, gravity, and magnetic fields are believed to collaborate or compete in the process of the formation of these observed structures \citep[e.g., ][]{2000ApJ...535..887K, 2007prpl.conf...63B, 2010A&A...518L.102A, 2012MNRAS.423.2680M, 2015ApJ...802...75K, 2020MNRAS.492..668B}. However, the relative importance of these three factors in the whole process remains to be unraveled. 

Several observations have presented evidence for a longitudinal flow of material along the filaments, leading to the idea that the filaments are long-lived but out-of-equilibrium flow structures that supply material to the central high-density hubs, the locations where the filaments converge to form an intricate and inhomogeneous network of filaments that in turn appear to accrete material from their surroundings \citep{2010A&A...520A..49S, 2013ApJ...766..115K, 2014ApJ...791..124G, 2014MNRAS.439.1996J, 2014A&A...561A..83P, 2016A&A...585A.158H, 2017A&A...607A..22R, 2018MNRAS.480.2939G, 2018ApJ...864..154D, 2018ApJ...855....9L, 2018A&A...613A..11W, 2019ApJ...875...24C, 2019A&A...623A..16S, 2019A&A...629A..81T}. Because these filaments are denser than their ambient medium, it is believed that some compression is essentially involved in their formation mechanism \citep{2013A&A...556A.153H}. This led to the suggestion that filamentary structures are formed by gravitational and/or turbulent compressive motions \citep[e.g., ][]{2001ApJ...553..227P, 2001ApJ...562..852H, 2002ApJ...576..870P, 2007prpl.conf...63B, 2011A&A...529L...6A}. The magnetic fields are expected to play a vital role by acting as agents that help to channel cloud accumulation and fragmentation \citep[e.g., ][]{1998ApJ...506..306N, 2000MNRAS.311..105F, 2006ApJ...647..997S, 2014ApJ...789...37V, 2015MNRAS.452.2410S}. Results from a study conducted by \cite{2018MNRAS.480.2939G} to investigate the structure of the magnetic field inside a self-gravitating filament formed in a turbulent environment suggest that the magnetic field geometry represents the flow pattern of the material inside and in the vicinity of the filaments. As cloud material tends to flow along the magnetic field lines, forming sheets or filaments, the low-density striations or low-column-density material tends to align parallel to the magnetic fields \citep{2013A&A...550A..38P, 2014arXiv1411.6096A, 2017MNRAS.472..647Z}. The gas around the filaments gets accreted onto them, making the field lines become oriented perpendicular to the filaments, which is often found in observations and simulations \citep{1992ApJ...399..108G, 2011ApJ...741...21C, 2011ApJ...734...63S, 2014ApJ...791..124G, 2018MNRAS.480.2939G}. But as the gas density increases, the flow pattern changes as an effect of the gravity-driven motion of cloud material. The longitudinal flow of material toward the clumps drags the field lines along with it, making the field lines parallel to the filaments \citep{2020NatAs...4.1195P}. Results from several studies have shown that the velocity and the magnetic fields are preferentially aligned \citep{2008PhRvL.100h5003M, 2009MNRAS.398.1082B, 2017A&A...604A..70I}. These results are consistent with the correlations found between the magnetic field direction and the density gradient \citep{2013ApJ...774..128S, 2013ApJ...775...77K, 2014ApJ...797...99K, 2016A&A...586A.138P, 2017A&A...607A...2S}. When magnetic field lines from opposite sides of the filament are connected, the resulting field lines become perpendicular at the internal vicinity of the filament \citep{2018MNRAS.480.2939G}.

Polarization observations in optical \citep[e.g., ][]{1976AJ.....81..958V, 1990ApJ...359..363G, 1992MNRAS.257...57B, 2008A&A...486L..13A, 2010ApJ...723..146F,2016A&A...588A..45N,2018MNRAS.476.4442N}, near-infrared \citep[e.g., ][]{1988MNRAS.230..321S,1995ApJ...448..748G,2010ApJ...716..299S,2011ASPC..449..207T,2015ApJ...803L..20S,2017ApJ...850..195E}, and submillimeter wavelengths \citep[e.g., ][]{2009ApJS..182..143M,2000ApJ...537L.135W,2014ApJ...794L..18Q,2016A&A...586A.138P, 2018ApJ...861...65S} have revealed the ubiquity of magnetic fields in the Galaxy. Studies that probed the relative orientation between the magnetic field and the elongated cloud structure suggest that they tend to be oriented either parallel or perpendicular to each other \citep{1990ApJ...359..363G, 1992ApJ...399..108G, 2011ApJ...741...21C, 2014prpl.conf..101L, 2011ApJ...734...63S}. The low-column-density structures are preferably oriented parallel to the local mean magnetic field, and the high-column-density structures are statistically oriented perpendicular to the same \citep{2011MNRAS.411.2067L, 2013MNRAS.436.3707L, 2013A&A...550A..38P, 2016MNRAS.461.3918H, 2016A&A...586A.138P, 2019MNRAS.485.2825A}. These findings are also consistent with the results obtained from numerical simulations \citep[e.g., ][]{1998ApJ...508L..99S, 1998ApJ...506..306N, 2008ApJ...687..354N, 2013A&A...556A.153H, 2014ApJ...789...37V, 2014ApJ...785...69C, 2015A&A...580A..49I, 2016MNRAS.459.1803W, 2016MNRAS.462.3602T}. While the polarization measurements in optical wavelengths can trace the field geometry in low-density inter-cloud media (ICM) and the periphery of molecular clouds, the polarized thermal emission due to the dust can trace the fields in the high-density parts of the cloud \citep[e.g., ][]{2009MNRAS.398..394W, 2009ApJ...704..891L, 2019ApJ...883....9S, 2020A&A...639A.133S}. Thus, by comparing the magnetic field geometry inferred from the optical and submillimeter wavelengths, the relationship between the magnetic field orientations inside the cloud and the ICM can be investigated.

The LDN 1172/1173/1174 \citep[hereafter L1172/1174; ][]{1962ApJS....7....1L} cloud complex was shown by \cite{2009ApJ...700.1609M} to be one of the typical examples of a hub-filament structure. The reflection nebula, NGC 7023, is illuminated by a Herbig Be (B2/3Ve) star, HD 200775 \citep{1994A&AS..104..315T,2006ApJ...653..657M}. The whole cloud complex forms the hub and a single filament of nearly 5 pc in length, which runs toward the southwest with respect to the Galactic plane, providing a \textquote{head-tail} appearance to the whole cloud \citep{2002A&A...385..909T}. Based on the \textit{Gaia} Data Release 2 (DR2) parallax measurements of the young stellar objects (YSOs) associated with the cloud, \cite{2020MNRAS.494.5851S} estimated a distance of 335$\pm$11 pc to L1172/1174. Thus, L1172/1174 forms a relatively nearby cloud composed of a single long filament terminating at a hub where low- and intermediate-mass star formation is currently active in L1174, which is located to the northwest of HD 200775, and in L1172, which is located on the filament \citep[e.g.,][]{1953AJ.....58...48W,2009ApJS..185..198K,2009ApJS..185..451K,2013AJ....145...35R,2013MNRAS.429..954Y}. Based on polarization measurements of more than 200 stars projected on the cloud in the R band, we determined the magnetic field geometry of the outer regions of the cloud complex. Apart from knowing the magnetic field orientation in the cloud complex, it is also interesting to study if there is any effect of HD 200775 on the ambient field orientation. Using the dust polarization in emission at submillimeter wavelengths obtained with the \textit{Planck} satellite, we made a low resolution magnetic field map of the low- and high-density regions of the cloud complex and compared them with those inferred from our R-band polarization measurements to understand the orientation of the field lines with respect to the column density distribution. This paper is organized in the following manner. We describe the details of our observations and data reduction in Sect. \ref{sec:obs_data}. The polarization results are presented in Sect. \ref{sec:result}, and we discuss our results in Sect. \ref{sec:dis}. Finally, we conclude our paper with a summary in Sect. \ref{sec:sum_con}.

\section{Observations and data reduction}\label{sec:obs_data}

\subsection{Polarization measurements in the R band}

Polarimetric observations of 42 fields (each field is a circle of 8$\arcmin$ in diameter) covering the cloud complex L1172/1174 were carried out using the ARIES Imaging POLarimeter \cite[AIMPOL;][]{2004BASI...32..159R}. AIMPOL is mounted as a back-end instrument at the Cassegrain focus of the 1.04m Sampurnanand Telescope, ARIES, Nainital, India. Only linear polarization can be obtained using AIMPOL, which consists of a half-wave plate (HWP) and a Wollaston prism. The HWP performs as a modulator and the Wollaston prism acts as a beam splitter, which splits the incoming light of each target into ordinary ($I_{o}$) and extraordinary ($I_{e}$) rays separated by $\sim$28 pixels \citep{2012MNRAS.419.2587E}. The CCD used during observations, is actually of 1024 $\times$ 1024 pixel$^{2}$, while frames for imaging polarimetry were obtained only within the central 325 $\times$ 325 pixel$^{2}$ area. The plate scale of the CCD is 1.48 arcsec pixel$^{-1}$. \citet{2004BASI...32..159R} provided a detailed description of AIMPOL. Each frame was obtained using a R-band filter ($\lambda_{eff}$ = 0.760 $\mu$m) by matching the Kron-Cousin passband. Table 2 in \citealp{2020MNRAS.494.5851S} provides the log of the optical R-band polarimetric observations in the upper section. After bias subtraction, flat correction of the images using the flux normalization formula from \cite{1998A&AS..128..369R}, we aligned and combined multiple images of an observed field.

We performed photometry of the selected pairs ($I_{o}$ and $I_{e}$ beams) of the observed sources using the Image Reduction and Analysis Facility (IRAF) DAOPHOT package to extract the fluxes of the $I_{o}$ and $I_{e}$ beams for individual sources. The selection of the $I_{o}$ and $I_{e}$ pair of each star from a given field is made automated using a program written in the Python language. The data reduction procedure is given in \cite{2013MNRAS.432.1502S,2015A&A...573A..34S,2017MNRAS.465..559S}. The details of our optical polarimetric observations were provided in our previous paper \citep{2020MNRAS.494.5851S}. The ratio R($\alpha$) is given by
\begin{equation}
R(\alpha) =\frac{\frac{I_{e}(\alpha)}{I_{o}(\alpha)}-1}{\frac{I_{e}(\alpha)}{I_{o}(\alpha)}+1} = Pcos(2\theta - 4\alpha),
\end{equation}
\noindent
where P is the fraction of total linearly polarized light and $\theta$ is the polarization angle of the plane of polarization. The $\alpha$ is the position of the fast axis of the HWP at 0$\degr$, 22.5$\degr$, 45$\degr$, and 67.5$\degr$ corresponding to four normalized Stokes parameters, q[$R(0\degr$)], u[$R(22.5\degr$)], q$_{1}$ [$R(45\degr$)], and u$_{1}$ [$R(67.5\degr$)], respectively. The polarization fraction P is estimated from $\sqrt{q^2 + u^2}$ and the polarization angle, $\theta$ = 0.5 tan$^{-1}$(u/q). Conventionally, $\theta$ is measured with respect to the celestial north-south axis (0$\degr$ toward the north celestial pole and increasing toward the east). The uncertainties in normalized Stokes parameters $\sigma_{R(\alpha)}$ ($\sigma_{q}$, $\sigma_{u}$, $\sigma_{q1}$ and $\sigma_{u1}$ in per cent) were estimated using the expression
\begin{equation}
\sigma_{R(\alpha)}=\frac{\sqrt{I_{e}+I_{o}+2I_{b}}}{I_{e}+I_{o}},
\end{equation}	
\noindent
where $I_{b}$[=$\frac{I_{be}+I_{bo}}{2}$] is the average background counts around the $I_{e}$ and $I_{o}$ beams of an observed source. The error in P\% and $\theta$ are estimated using,
\begin{equation}
\sigma_{p} =\frac{1}{P}\times\sqrt{q^{2}\sigma_{q}^{2}+u^{2}\sigma_{u}^{2}}, \hspace{0.7cm}\sigma_{\theta} = \frac{1}{2P^{2}}\times\sqrt{q^{2}\sigma_{u}^{2}+u^{2}\sigma_{q}^{2}}\hspace{0.2cm}rad.
\end{equation}
	
\noindent
In order to determine the reference direction of the polarizer, we observed six polarized standard stars (HD 236633, BD+59$^{\circ}$389, HD 19820, HD 204827, HD 25443, HD 15445) from the list given by \cite{1992AJ....104.1563S} during each observing run. Table \ref{tab:pol_std} provides the optical R-band polarimetric results of the six polarimetric standard stars. HD204827 was observed in a single run. Mean degrees of polarization (P\%) of HD 236633, BD+59$^{\circ}$389, HD 19820, HD 25443, HD 15445 are 5.0, 6.1, 4.4, 4.8, 3.4 with standard deviations 0.2, 0.3, 0.2, 0.1, 0.2, respectively. Therefore, the estimated P\% of the observed standard stars are found to be comparable as provided by \cite{1992AJ....104.1563S}. The difference between the computed $\theta$ of individual standard star and its value given in \cite{1992AJ....104.1563S} was used to estimate the offset in the polarizer. Since 2004, the instrumental polarization of AIMPOL has been monitored by observing unpolarized standard stars, which is found to be stable \citep[typically less than $\sim$ 0.1\%; see][]{2004BASI...32..159R, 2008MNRAS.388..105M, 2011MNRAS.411.1418E, 2014Ap&SS.350..251S, 2016A&A...588A..45N, 2018MNRAS.476.4782S}. Corrections for instrumental polarization were made for our polarimetric results.

\begin{table}
\centering
\caption{Polarized standard stars observed in the R band ($\lambda_{eff}$=0.760 $\mu$m).}
	\footnotesize
	\label{tab:pol_std}
	\begin{tabular}{ccc} 
		\hline
		\hspace*{-6.5cm}Observing date & \hspace*{-5.0cm}P $\pm~\sigma_{P}$ & \hspace*{1.8cm}$\theta\pm\sigma_{\theta}$\\
		\hspace*{-6.5cm} & \hspace*{-5.0cm}(\%) & \hspace*{1.8cm}($^{\circ}$)\\
        \hline
        \multicolumn{3}{c}{\textbf{HD 236633} ($^{a}$Standard values: 5.38 $\pm$ 0.02\%, 93.$^{\circ}$04 $\pm$ 0.$^{\circ}$15)}\\
        \hline
        \hspace*{-6.5cm}11 Oct. 2015 & \hspace*{-5.0cm}5.4$\pm$0.3 &\hspace*{1.5cm}97$\pm$2\\
         \hspace*{-6.5cm}3 Nov. 2015 & \hspace*{-5.0cm}4.9$\pm$0.1 & \hspace*{1.5cm}100$\pm$1\\
         \hspace*{-6.5cm}15 Nov. 2015 & \hspace*{-5.0cm}4.9$\pm$0.1 & \hspace*{1.5cm}101$\pm$1\\
         \hspace*{-6.5cm}16 Nov. 2015 & \hspace*{-5.0cm}5.1$\pm$0.1 & \hspace*{1.5cm}102$\pm$1\\
         \hspace*{-6.5cm}17 Nov. 2015 & \hspace*{-5.0cm}4.9$\pm$0.1 & \hspace*{1.5cm}101$\pm$1\\
         \hspace*{-6.5cm}15 Dec. 2015 & \hspace*{-5.0cm}4.8$\pm$0.1 & \hspace*{1.5cm}101$\pm$1\\
         \hspace*{-6.5cm}23 Oct. 2016 & \hspace*{-5.0cm}5.2$\pm$0.1 & \hspace*{1.5cm}100$\pm$1\\
         \hspace*{-6.5cm}25 Oct. 2016 & \hspace*{-5.0cm}5.0$\pm$0.1 & \hspace*{1.5cm}99$\pm$1\\
         \hspace*{-6.5cm}26 Oct. 2016 & \hspace*{-5.0cm}4.8$\pm$0.1 & \hspace*{1.5cm}98$\pm$1\\
         \hspace*{-6.5cm}27 Oct. 2016 & \hspace*{-5.0cm}4.9$\pm$0.1 & \hspace*{1.5cm}100$\pm$1\\
         \hspace*{-6.5cm}13 Oct. 2017 & \hspace*{-5.0cm}4.9$\pm$0.2 & \hspace*{1.5cm}102$\pm$1\\
         \hspace*{-6.5cm}19 Oct. 2017 & \hspace*{-5.0cm}5.1$\pm$0.2 & \hspace*{1.5cm}100$\pm$1\\
         \hspace*{-6.5cm}20 Oct. 2017 & \hspace*{-5.0cm}5.0$\pm$0.2 & \hspace*{1.5cm}101$\pm$1\\
         \hspace*{-6.5cm}21 Oct. 2017 & \hspace*{-5.0cm}5.3$\pm$0.2 & \hspace*{1.5cm}102$\pm$1\\
         \hspace*{-6.5cm}26 Oct. 2017 & \hspace*{-5.0cm}5.1$\pm$0.2 & \hspace*{1.5cm}101$\pm$1\\
         \hspace*{-6.5cm}27 Oct. 2017 & \hspace*{-5.0cm}4.9$\pm$0.1 & \hspace*{1.5cm}100$\pm$1\\
         \hline
         \multicolumn{3}{c}{\textbf{BD+59$^{\circ}$389} ($^{a}$Standard values: 6.43 $\pm$ 0.02 \%, 98.$^{\circ}$14 $\pm$ 0.$^{\circ}$10)}\\
        \hline
         \hspace*{-6.5cm}11 Oct. 2015 & \hspace*{-5.0cm}6.2$\pm$0.2 & \hspace*{1.5cm}104$\pm$1\\
         \hspace*{-6.5cm}2 Nov. 2015 & \hspace*{-5.0cm}6.4$\pm$0.1 & \hspace*{1.5cm}106$\pm$1\\
         \hspace*{-6.5cm}3 Nov. 2015 & \hspace*{-5.0cm}6.3$\pm$0.1 & \hspace*{1.5cm}105$\pm$1\\
         \hspace*{-6.5cm}15 Nov. 2015 & \hspace*{-5.0cm}6.0$\pm$0.1 & \hspace*{1.5cm}106$\pm$1\\
         \hspace*{-6.5cm}16 Nov. 2015 & \hspace*{-5.0cm}6.4$\pm$0.1 &\hspace*{1.5cm}106$\pm$1\\
         \hspace*{-6.5cm}17 Nov. 2015 & \hspace*{-5.0cm}6.2$\pm$0.1 & \hspace*{1.5cm}106$\pm$1\\
         \hspace*{-6.5cm}25 Oct. 2016 & \hspace*{-5.0cm}6.0$\pm$0.1 & \hspace*{1.5cm}104$\pm$1\\
         \hspace*{-6.5cm}28 Oct. 2016 & \hspace*{-5.0cm}5.6$\pm$0.1 & \hspace*{1.5cm}105$\pm$1\\
         \hspace*{-6.5cm}22 Nov. 2016 & \hspace*{-5.0cm}6.3$\pm$0.1 & \hspace*{1.5cm}106$\pm$1\\
         \hspace*{-6.5cm}27 Nov. 2016 & \hspace*{-5.0cm}6.4$\pm$0.3 & \hspace*{1.5cm}109$\pm$1\\
         \hspace*{-6.5cm}13 Oct. 2017 & \hspace*{-5.0cm}5.9$\pm$0.1 & \hspace*{1.5cm}106$\pm$1\\
         \hspace*{-6.5cm}14 Oct. 2017 & \hspace*{-5.0cm}6.4$\pm$0.1 & \hspace*{1.5cm}107$\pm$1\\
         \hspace*{-6.5cm}17 Oct. 2017 & \hspace*{-5.0cm}5.7$\pm$0.2 & \hspace*{1.5cm}106$\pm$1\\
         \hspace*{-6.5cm}18 Oct. 2017 & \hspace*{-5.0cm}5.7$\pm$0.2 & \hspace*{1.5cm}107$\pm$1\\
         \hspace*{-6.5cm}19 Oct. 2017 & \hspace*{-5.0cm}6.0$\pm$0.2 & \hspace*{1.5cm}107$\pm$1\\
         \hspace*{-6.5cm}20 Oct. 2017 & \hspace*{-5.0cm}6.1$\pm$0.2 & \hspace*{1.5cm}107$\pm$1\\
         \hspace*{-6.5cm}21 Oct. 2017 & \hspace*{-5.0cm}6.9$\pm$0.2 & \hspace*{1.5cm}105$\pm$1\\
         \hspace*{-6.5cm}26 Oct. 2017 & \hspace*{-5.0cm}6.0$\pm$0.2 & \hspace*{1.5cm}103$\pm$1\\
         \hline
         \multicolumn{3}{c}{\textbf{HD 19820} ($^{a}$Standard values: 4.526 $\pm$ 0.025 \%, 114.$^{\circ}$46 $\pm$ 0.$^{\circ}$16)}\\
        \hline
		\hspace*{-6.5cm}15 Dec. 2015 & \hspace*{-5.0cm}4.6$\pm$0.1 & \hspace*{1.5cm}123$\pm$1\\
        \hspace*{-6.5cm}22 Nov. 2016 & \hspace*{-5.0cm}4.3$\pm$0.1 & \hspace*{1.5cm}126$\pm$1\\
        \hspace*{-6.5cm}26 Nov. 2016 & \hspace*{-5.0cm}4.2$\pm$0.1 & \hspace*{1.5cm}124$\pm$1\\
        \hline
        \textbf{HD 204827} ($^{a}$Standard values: 4.893 $\pm$ 0.029 \%, 59.$^{\circ}$10 $\pm$ 0.$^{\circ}$17)\\
        \hline
        \hspace*{-6.5cm}23 Oct. 2016 & \hspace*{-5.0cm}4.7$\pm$0.1 & \hspace*{1.5cm}66$\pm$1\\
        \hline
        \textbf{HD 25443} ($^{a}$Standard values: 4.734 $\pm$ 0.045 \%, 133.$^{\circ}$65 $\pm$ 0.$^{\circ}$28)\\
        \hline
        \hspace*{-6.5cm}26 Nov. 2016 & \hspace*{-5.0cm}4.7$\pm$0.2 & \hspace*{1.5cm}143$\pm$1\\
        \hspace*{-6.5cm}27 Nov. 2016 & \hspace*{-5.0cm}4.8$\pm$0.2 & \hspace*{1.5cm}143$\pm$1\\
        \hline
        \textbf{HD 15445} ($^{a}$Standard values: 3.683 $\pm$ 0.072 \%, 88.$^{\circ}$91 $\pm$ 0.$^{\circ}$56)\\
        \hline
        \hspace*{-6.5cm}22 May 2017 & \hspace*{-5.0cm}3.6$\pm$0.2 & \hspace*{1.5cm}107$\pm$2\\
        \hspace*{-6.5cm}23 May 2017 & \hspace*{-5.0cm}3.2$\pm$0.2 & \hspace*{1.5cm}102$\pm$2\\
        \hline
	\end{tabular}\\
    {$^{a}$ Values in the R band from \cite{1992AJ....104.1563S}.}
\end{table}
\begin{table}
	\begin{center}
	\small	
	\caption{Polarimetric results of the YSO candidates.}\label{tab:yso_pol_results_table}
	\begin{tabular}{lcccc}\hline
		ID & $l$  & $b$ & \pr$\pm$$\sigma$\pr & \tr$\pm$$\sigma$\tr\\
		 & $(\degr)$ & $(\degr)$ & (\%) & $(\degr)$\\
		 \hline
		3 & 104.041838 & 14.259696 & 1.4$\pm$0.6 & 156$\pm$10\\
		13/13b$^{*}$ & 104.029883 & 14.063977 & 2.6$\pm$0.3 & 182$\pm$3\\
		18 & 104.372912 & 14.192269 & 2.6$\pm$0.6 & 202$\pm$6\\
		27 & 104.036152 & 14.301923 & 2.9$\pm$0.7 & 181$\pm$7\\
        \hline
	\end{tabular}
	\end{center}
$^{*}$ This source appeared as double sources in \textit{Gaia} DR2 (LkH$\alpha$ 428 N/S; \citealp{2009ApJS..185..451K}).
\end{table}

\subsection{\textit{Planck} polarization measurements in submillimeter}

\textit{Planck} observed the whole sky in nine frequency bands (30-857 GHz) in total intensity, and up to 353 GHz in polarization \citep{2014A&A...571A...1P}. The data were thus used to produce the first all-sky map of the polarized emission from dust at submillimeter wavelengths \citep{2016A&A...594A...1P}. We used the intensity and polarization data only at 353 GHz as this is the highest frequency channel with polarization capabilities and the one with best signal-to-noise ratio (S/N) for dust polarization \citep{2015A&A...576A.104P}. We used the whole sky map (bandpass leakage corrected) at 353 GHz provided by \textit{Planck} Legacy Archive\footnote{\url{http://www.cosmos.esa.int/web/planck/pla/}}. The polarization of the cosmic microwave background (CMB) has a negligible contribution to the sky polarization toward the molecular clouds at 353 GHz \citep{2016A&A...586A.133P}. So, CMB polarization was not taken into account in our analysis. The I, Q, and U maps analyzed here have been constructed using the gnomonic projection of the HEALPix\footnote{\url{http://healpix.sourceforge.net}} \citep{2005ApJ...622..759G} all-sky maps. We used {\it healpy} \citep{Zonca2019} to extract and analyze data for calculation of P\% and $\theta$. {\it healpy} is a Python package to handle pixelated data on the sphere.

We estimated the Stokes I, Q, and U parameters from the smoothed \textit{Planck} map of a 8$\degr$ square area obtained from the \textit{Planck} 353 GHz image centered at HD 200775. The Stokes parameter maps are shown in accordance with the IAU convention (i.e., the polarization angle $\psi$ = 0$\arcdeg$ toward the Galactic north, increasing toward the Galactic east) \citep{refId0}. The angle of the magnetic field projected on the sky plane can be obtained by adding 90$\arcdeg$ to the polarization angle ($\psi$ + 90$\arcdeg$). At 353 GHz, the \textit{Planck} data have an angular resolution of 4.8$\arcmin$. 

\subsection[12CO (1-0) molecular line observations using TRAO]{$^{12}$CO (1-0) molecular line observations using TRAO}

As a part of a comprehensive study of L1172/1174 to understand the gas dynamics (Sharma et al. 2021, under preparation), the whole cloud of L1172/L1174 was observed in $^{12}$CO, C$^{18}$O, N$_{2}$H$^{+}$ (1-0), and CS (2-1) transitions using the on-the-fly (OTF) mapping technique and the 14 m diameter single-dish telescope of Taedeuk Radio Astronomy Observatory (TRAO) in Daejeon, South Korea, between November 16-28, 2018. Here we present only the $^{12}$CO line results mainly to estimate the magnetic field strengths in the cloud using the $^{12}$CO linewidths. The back-end system with fast fourier transform spectrometer has 4096 $\times$ 2 channels at 15 kHz resolution ($\sim$0.05 km~s$^{-1}$ at 110 GHz). The typical rms noise in one channel was $\sim$0.35 K for $^{12}$CO lines in T$_{A}^{*}$ scale. At 115 GHz the beam size (HPBW) of the telescope is about $47^{"}$ and the fraction of the beam pattern subtending main beam (beam efficiency) is 41$\pm$2\% \citep{2019JKAS...52..227J}. The system temperature was 550 K-600 K during the observations. The spectra were reduced using CLASS software of the IRAM GILDAS software package.

\subsection{Archival \textit{Gaia} DR2}

\textit{Gaia} DR2 presents positions, parallaxes, and proper motions of more than a billion objects \citep{2018A&A...616A...1G} with unprecedented precision. Typical uncertainties in parallax measurements of sources brighter than $\sim$14 mag are around 0.4 mas. Sources having $G$-mag around 17 have typical uncertainties $\sim$0.1 mas and $\sim$0.7 mas for faint sources ($G$-mag around 20) \citep{2018A&A...616A...9L}. But if the relative uncertainties in parallax values were $\gtrsim$ 20\%, the corresponding distances would not follow the simple inversion of their parallaxes \citep{2015PASP..127..994B}. Recently, \cite{2018AJ....156...58B} provided a probabilistic estimate of the stellar distances from the parallax measurements in \cite{2018A&A...616A...1G}, using an exponentially decreasing space density prior that is based on a galactic model. In our analysis, the stellar distances were obtained from \cite{2018AJ....156...58B}, by giving a search around a circle of radius of 1$\arcsec$ around the source positions. 

\section{Results} \label{sec:result}
\begin{figure*}
    \includegraphics[height=11cm, width=10cm]{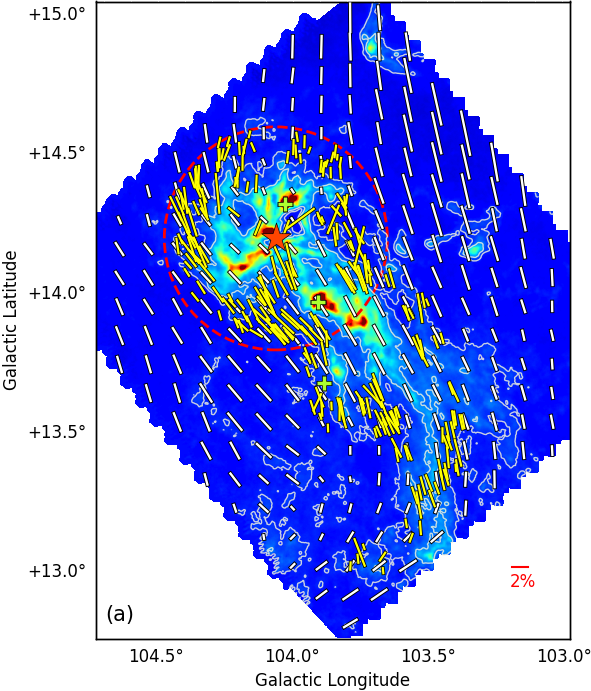}
    \includegraphics[width=8.2cm, height=11.5cm]{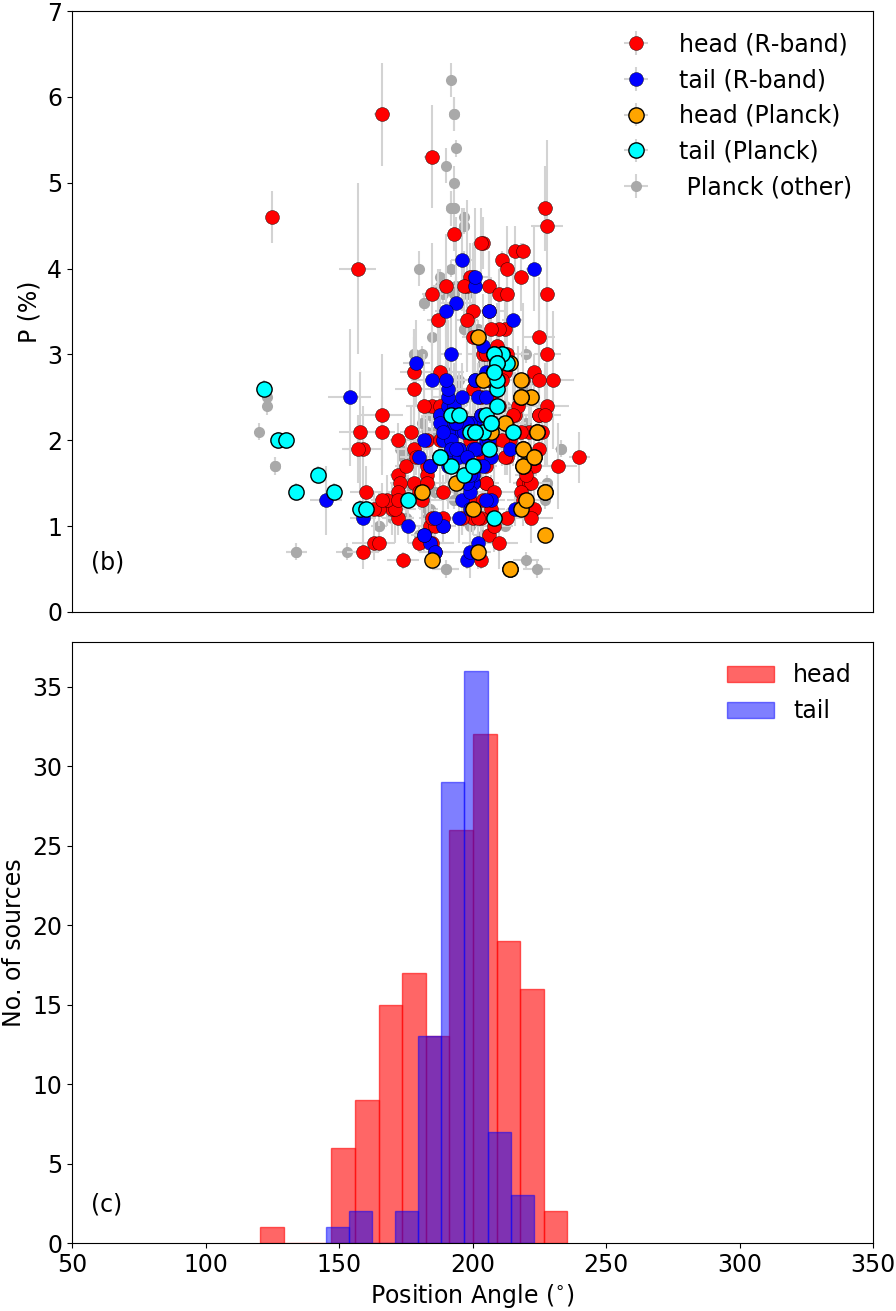}
    \caption{Polarimetric results obtained toward the L1172/1174 cloud complex. \textbf{(a):} R-band (yellow) and \textit{Planck} (white) polarization vectors (90$\degr$ rotated) overlaid on the \textit{Herschel} dust column density map of L1172/1174 obtained from \cite{2010A&A...518L.102A}. The contours are drawn at 9$\times$10$^{20}$ cm$^{-2}$ and 15$\times$10$^{20}$ cm$^{-2}$ levels to reveal the cloud structure. HD 200775 is marked with a red star symbol. The dashed red circle represents the head region of the cloud complex. The green colored plus symbols show the locations of L1173 (south), L1172 (middle), and L1174 (north). The lengths and the orientations of the vectors correspond to the degree of polarization and the position angles measured with respect to the Galactic north increasing eastward, respectively. A polarization vector (red) corresponding to 2\% and oriented at 90$\degr$ is shown for reference. \textbf{(b):} Plot of degree of polarization versus position angle values for the 249 sources observed by us in the R band and \textit{Planck}. The R-band polarization vectors from the head and the tail regions are shown using solid red and blue circles, respectively. Similarly, for \textit{Planck}, the measurements from the head and the tail regions are shown using solid orange and cyan circles, respectively. The \textit{Planck} polarization measurements from the ICM are shown with solid gray circles. \textbf{(c):} Histograms of the \tr belonging to the head and the tail regions of L1172/1174 obtained from our R-band polarization measurements with the bin size of $\sim$9$\degr$.} \label{fig:cepheus_pol_YSO}
\end{figure*}

\subsection{Optical and submillimeter polarization}

Although the Stokes parameters Q and U can be both positive or negative, the polarization P derived from these parameters is always positive and thus it has a positive bias, especially for sources with low S/Ns. In order to eliminate this bias, we estimated the debiased P using P=$\sqrt{P^{2}-\sigma_{P}^{2}}$, where $\sigma_{P}$ is the error in P \citep{1974ApJ...194..249W}. There are 301 sources for which the ratios of the P\%, \tp, and distance to their corresponding errors, P/$\sigma_{P}$, \tp/$\sigma_{\theta_{P}}$, and d/$\sigma_{d}$, respectively, are greater than three. Of the 301 sources, we included 249 of them, which show uncertainty in polarization measurement $\lesssim$ 1\%. This corresponds to a \textit{Gaia} $G$ magnitude brighter than $\sim17$. Though we have polarimetric results for four YSO candidates located toward L1172/1174, these sources are not included in our analysis as there could be additional effects on polarization measurements due to possible presence of circumstellar dust \citep[e.g.,][]{1977IAUS...75..179S, 1978A&A....70L...3E, 1979ApJ...229L.137B, 1982A&AS...48..153B, 1985ApJS...59..277B, 1988PThPS..96...37S, 2005ASPC..343..128M, 2005MNRAS.359.1049V}. These sources are shown separately in Table \ref{tab:yso_pol_results_table}. The YSO candidates with IDs 3, 13/13b$^{*}$, and 18 are detected in \textit{Gaia} DR2 and hence were listed in Table 3 of \cite{2020MNRAS.494.5851S}. The last source (ID 27) has no detection in \textit{Gaia} DR2, and therefore it was listed in Table 7 of \cite{2020MNRAS.494.5851S}. In Fig. \ref{fig:cepheus_pol_YSO} (a) we show the results from our R-band polarization measurements (lines shown in yellow) of 249 stars. Table \ref{tab:res249} in the appendix lists the polarimetric measurements of these stars. The lengths and orientations of the polarization vectors correspond to the degree of polarization (P$_\mathrm{R}$) in per cent and the position angle ($\theta_\mathrm{R}$) in degree measured from the north to east, respectively. The polarization vectors are plotted over the $\sim$2$\degr$ hydrogen column density map made using the images obtained with the \textit{Herschel} satellite \citep{2010A&A...518L.102A}. The median value of the column density toward the 249 stars observed by us is found to be 9$\times$10$^{20}$ cm$^{-2}$. The outermost contour in Fig. \ref{fig:cepheus_pol_YSO} (a) is drawn at this level, which corresponds to an extinction of $\sim$1 magnitude converted using the relationship between the column density and the extinction derived by \citet{1978ApJ...224..132B}. 

Interstellar polarization due to the differential extinction of starlight by the asymmetric dust grains aligned with the ambient magnetic field was reported more than half a century ago \citep{1949Sci...109..166H, 1949ApJ...109..471H}. For years, a number of grain alignment paradigms have been proposed. As the dust grains are composed of paramagnetic material, they contain unpaired electrons. Through the Barnett effect they can develop an internal magnetization \citep{1976Ap&SS..43..291D}. Due to the presence of magnetic field, the directions of angular momentum of these magnetized dust grains are aligned along the direction of field lines via the continuing radiative torques over the period of Larmor precession. According to \cite{1951ApJ...114..206D}, on account of paramagnetic loses by dissipation of the angular momentum components perpendicular to magnetic field direction, the dust grains get aligned with field lines. Another mechanism was proposed based on mechanical grain alignment \citep{1952Natur.169..322G, 1952MNRAS.112..215G}. However, a majority of theoretical and observational studies indicated the mechanism of radiative alignment torque (RAT), which became the best to explain the interstellar polarization. This mechanism was first introduced by \cite{1976Ap&SS..43..291D} and later studied by \cite{1996ApJ...470..551D, 1997ApJ...480..633D}. \cite{2007MNRAS.378..910L} formulated the analytical model of RAT paradigm. According to RAT theory, the aspherical dust grains rotate due to radiative torque and orient with their long axes perpendicular to magnetic field lines\citep[B-RAT;][]{2007MNRAS.378..910L}. In another scenario of RAT mechanism, \textit{\textbf{k}}-RAT (significant in the vicinities of bright stars), the dust grains precess about the direction of radiation \citep{2007MNRAS.378..910L, 2017ApJ...839...56T}. The selective extinction due to the aligned, elongated dust grains builds the polarization vectors oriented along the direction of the plane-of-sky component of the magnetic field (\bpos\!\!). Now, the thermal continuum emission would be polarized along the long axes of the elongated dust grains, which is perpendicular to the field direction. Therefore, the polarization vectors deduced by submillimeter emission have to be rotated by 90$\degr$ to estimate the geometry of the \bpos \citep{1996ASPC...97..325G, 2003ApJ...592..233W}.

In Fig. \ref{fig:cepheus_pol_YSO} (a), the \textit{Planck} polarization vectors are presented, which are 90$\degr$ rotated by their original orientations to indicate the magnetic field directions. Here again, the lengths and orientations of the polarization vectors correspond to the degree of polarization (P$_\mathrm{p}$) and the position angle ($\theta_\mathrm{P}$) in degree measured from the north toward the east, respectively. For the purpose of analysis, R-band polarization vectors lying within a circular region of 0.4$\degr$ (the extent over which we have R-band polarization across the cloud complex) radius around HD 200775 are considered as part of the head and those lying outside of this region are considered as part of the tail region. The \pr and the \tr for the sources lying toward the head and the tail regions are shown in Fig. \ref{fig:cepheus_pol_YSO} (b). The results of \pr and \tr in both head and tail regions are presented in Table \ref{tab:pol_results_table}. The median absolute deviations (MADs), more resilient to outliers in a data set, of \pr and \tr are also listed as uncertainties. The histograms of \tr from the head and the tail regions are shown in Fig. \ref{fig:cepheus_pol_YSO} (c).

The \textit{Planck} polarization measurements, \pp and \tp, for the head and the tail regions are also shown in Fig. \ref{fig:cepheus_pol_YSO} (b). Here, the \textit{Planck} polarization vectors that fall within the 9$\times$10$^{20}$ cm$^{-2}$ contour are considered as associated with the cloud complex and those lying outside (but within the \textit{Herschel} field) are considered as polarization from the ICM. Of the measurements falling within the 9$\times$10$^{20}$ cm$^{-2}$ contour, those lying within the circular region of 0.4$\degr$ radius are considered as the part of the head region while those falling outside the circular region of 0.4$\degr$ radius are considered as the part of the tail. The number of data points is lesser compared to the R-band polarization due to the coarse resolution of the \textit{Planck} measurements. We used the threshold of P/$\sigma_{P}$>3 in the analysis of \textit{Planck} data as well. The \textit{Planck} polarization measurements were debiased using the equations given by \cite{2015A&A...574A.135M}. The uncertainties of \pp and \tp are estimated using the equations provided by \cite{2015A&A...574A.135M}. Typical uncertainties in the measurements of \pp and \tp are found to be 0.1\% and 1$\degr$, respectively. The results of \pp and \tp in both head and tail regions of the cloud complex and also in the ICM are presented in Table \ref{tab:pol_results_table}. 

\begin{table}
	\centering
	\caption{Polarimetric results in head and tail of L1172/1174 and ICM.}\label{tab:pol_results_table}
	\begin{tabular}{ccccc}\hline
	    &\multicolumn{2}{c}{Range} &\multicolumn{2}{c}{Median}\\
		Region & P (\%) & $\theta (\degr)$ & P (\%) & $\theta (\degr)$\\
		 \hline
		\multicolumn{5}{c}{Optical R-band polarimetric results}\\\\
		head & 0.6$-$5.8 & 125$-$240 & 2.1$\pm$0.7 & 203$\pm$15\\
        tail & 0.6$-$4.1 & 145$-$223 & 2.0$\pm$0.4 & 196$\pm$6\\
        \hline   
        \multicolumn{5}{c}{\textit{Planck} polarimetric results}\\\\
        head & 0.5$-$3.2 & 181$-$227 & 1.8$\pm$0.6 & 218$\pm$6\\
        tail & 1.1$-$3.0 & 122$-$215 & 2.1$\pm$0.5 & 201$\pm$8\\
        ICM & 0.5$-$6.2 & 120$-$233 & 2.2$\pm$0.7 & 195$\pm$11\\
        \hline
	\end{tabular}
\end{table} 

\subsection{$^{12}$CO gas distribution in L1172/1174 complex}

The $^{12}$CO velocity for the entire cloud complex ranges from -2.8 km~s$^{-1}$ to 7.8 km~s$^{-1}$. The average full width at half maximum (FWHM) of the $^{12}$CO is found to be $\sim$2.0 km~s$^{-1}$. The cavity surrounding HD 200775 is conspicuous with high intensity peaks located along the rim of the cavity. The $^{12}$CO is found to show gas structures, especially to the east and west of HD 200775, where the gas components in the velocity range of -2.5 km~s$^{-1}$ to 0 km~s$^{-1}$ and 5.0 km~s$^{-1}$ to 7.5 km~s$^{-1}$, respectively are located, supporting the earlier reporting of the presence of bipolar outflow lobes \citep{1986A&A...163..194W}. The V$_{lsr}=$2.65 km~s$^{-1}$ of the cloud complex is determined from the N$_{2}$H$^{+}$ line detected in the cores (Sharma et al. 2021, under preparation). The V$_{lsr}$ is computed as a mean of L1174 (2.5 km~s$^{-1}$) and L1172 (2.8 km~s$^{-1}$) regions. Multiple velocity components in the $^{12}$CO line profiles are found in different parts of the cloud complex. The gas dynamics of the region surrounding HD 200775 will be presented in a subsequent article (Sharma et al. 2021, under preparation). In this work, we used the $^{12}$CO linewidths to calculate the strength of the \bpos\!\!.

\section{Discussion}\label{sec:dis}

\subsection{Magnetic field geometry of the L1172/1174 complex}

Interstellar dust grains produce extinction of light from the background stars as well as emit thermal radiation. When the unpolarized starlight impinges on a series of aspherical dust grains, the electromagnetic wave is absorbed maximum along the long axis of grain. The transmitted radiation is parallel to the short axis and becomes partially plane polarized, oriented along the \bpos\!\! \citep{2005ASPC..343..321W}. On the other hand, thermal emission from the dust grains is maximum along the longer axis, thus becomes polarized perpendicular to \bpos\!\!. Since dust grains are coupled with the magnetic field, measurements of dust polarization provide information on the structure and the strength of the magnetic field. The value of position angle depends on the orientation of the \bpos and the efficiency of the dust grain alignment with the field. It also depends on the variation in the orientation of \bpos along the line-of-sight \citep{1985ApJ...290..211L}. The polarization measurements in the optical bands trace the \bpos only in the outskirts of the molecular clouds where the extinction (A$_{V}$) is relatively low \citep{1989AJ.....98..611G, 1999A&A...349..912H}. On the other hand, the polarization measurements in submillimeter or millimeter can infer the \bpos inside the cloud where A$_{V}$ is relatively high \citep[e.g., ][]{1984ApJ...284L..51H, 1995ASPC...73...45G, 1999A&A...344..668G}.
\begin{figure*}
\centering
	\includegraphics[width=\textwidth, height=5.5cm]{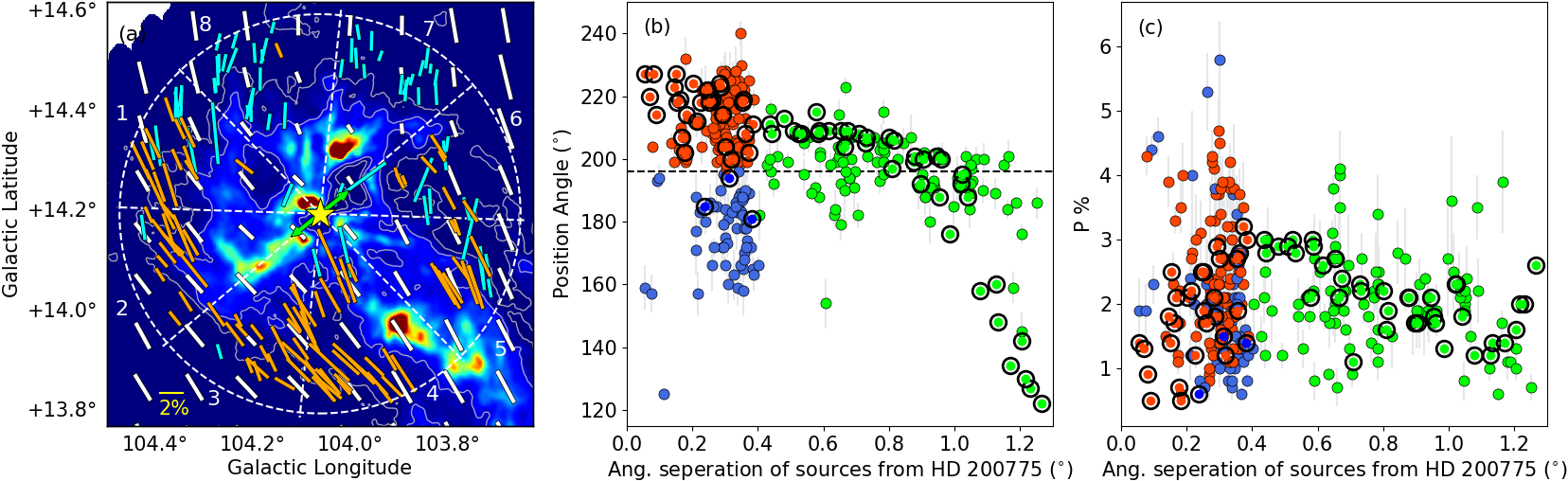}
	\caption{Variation in the projected magnetic field geometry toward the L1172/1174 cloud complex. \textbf{(a)} Optical R-band and \textit{Planck} polarimetric results overplotted on \textit{Herschel} column density map. Location of HD 200775 (yellow star) is also shown. The green colored arrows imply the directions of outflow from HD 200775. The eight sectors are indicated using dashed white lines and also marked. Orange lines indicate the polarization measurements in the R band with $\theta_{R} \geqslant$ 196$\degr$ and cyan lines represent the same with $\theta_{R} <$ 196$\degr$. White lines represent \textit{Planck} polarimetric results with 90$\degr$ rotation. A polarization vector corresponding to 2\% is shown for reference. \textbf{(b)} Variation in the $\theta_{R}$ of the sources with respect to the angular separation from HD 200775. Upto 0.4$\degr$ (sections 1-8) there are sources located toward the head region with two sets of distribution of position angles $\sim$211$\degr$ (red filled circles, shown in orange lines in \textbf{(a)}) and $\sim$179$\degr$ (blue filled circles, shown in cyan lines in \textbf{(a)}). The green filled circles are the sources distributed toward the tail region. The dashed horizontal line indicates the mean $\theta_{R}$ value ($\sim 196\degr$) of the same sources. Position angles obtained from \textit{Planck} observations are shown using thick open black circles (shown in white lines in \textbf{(a)}). \textbf{(c)} Variation in the P$_{R}$\% of the sources with respect to the angular separation from HD 200775. The symbols represent the same as described in panel \textbf{(b)}.}
	\label{fig:p_pa_sec_d}
\end{figure*}

The measured polarization in optical wavelengths is caused by the dust grains that are lying all along the light of sight (within the pencil beam) between the star and us. Though the dust grains present within the cloud are mostly responsible for the observed polarization, those that are present foreground to the cloud can also contribute. To infer the amount of polarization caused by the dust grains within the cloud (thus the magnetic field geometry), it is necessary to remove the foreground component. Generally, the stars located foreground to the cloud are used to estimate the foreground component of the polarization. The \pr versus distance and \tr versus distance plots of the observed sources are shown in Fig. 3 (a) and (b) in \citep{2020MNRAS.494.5851S}, respectively, using solid gray triangles (in equitorial coordinate system). The sources having $G$ magnitudes fainter than $\sim$ 17 were also included in that figure. Among the stars observed by us, only one star, located at 306 pc, is foreground to the cloud. Therefore, we searched in \citet{2000AJ....119..923H} catalog to get more foreground sources having polarization measurements. The search was made within a circular region of 6$\degr$ radius about HD 200775. We obtained a total of 26 sources around HD 200775. In our analysis, we discarded four sources, HD 200775, HD 193533, HD 208947 and HD 203467 (reason for not considering these sources was explained in \cite{2020MNRAS.494.5851S}). Of the remaining 22, we obtained the \textit{Gaia} DR2 distance for 17 sources from the \citet{2018AJ....156...58B} catalog. 

The degree of polarization (P$_\mathrm{H}$) in per cent and polarization position angles ($\theta_\mathrm{H}$) in degree of the 17 sources are also shown using filled circles in black in both (a) and (b) of Fig. 3 of \cite{2020MNRAS.494.5851S}, respectively. As the foreground sources show significantly low degree of polarization, we set no constraints on the P/$\sigma_{P}$ values during their selection. Of the 17 sources, 13 sources are found to be located at distances less than 335 pc (distance of L1172/1174) and four sources are at distances greater than 335 pc. The mean values of \ph and \th for the 13 foreground sources are found to be 0.1\% and 122$\degr$, respectively and the corresponding mean values for the four background sources are found to be 1.9\% and 199$\degr$, respectively. Both \ph and \th values of the four background sources are found to be consistent with the values of the sources observed by us toward L1172/1174. Using the mean values of \ph and \th obtained for the 13 stars, we calculated the mean Stokes parameters Q$_{\mathrm{fg}}$(= Pcos2$\theta$) and U$_{\mathrm{fg}}$(= Psin2$\theta$) as -0.043 and -0.089, respectively. We subtracted these values from the Stokes parameters of the observed stars vectorially to calculate the foreground-corrected percentage of polarization (P$_{\mathrm{c}}$) and the position angle ($\theta_\mathrm{C}$) values. No significant changes are noticed in the results after correcting for the foreground polarization. As a result, the R-band polarization vectors (in yellow) presented in Fig. \ref{fig:cepheus_pol_YSO} (a) represent the B$_\mathrm{POS}$ geometry toward the periphery of L1172/1174 complex. 

As evident from Fig. \ref{fig:cepheus_pol_YSO} (a) and (b), the orientations of \bpos obtained from the R band (\tr=203$\degr$ for the head and \tr=196$\degr$ for the tail) and \textit{Planck} (\tp=212$\degr$ for the head and \tp=200$\degr$ for the tail) observations are found to be in good agreement all along the structure of the cloud complex. Such correlations between the magnetic field geometries inferred from the optical and the \textit{Planck} polarization measurements are reported in a number of studies \citep[e.g., ][]{2016A&A...596A..93S, 2016A&A...586A.138P, 2019ApJ...871L..15G}. This implies that the magnetic fields inferred from the \textit{Planck} polarization and from the R-band polarization measurements toward the high-density parts and toward the outer envelope of L1172/1174, respectively, are well correlated with each other. Though consistent with the \textit{Planck} measurements, small-scale substructures in the magnetic field toward the envelope region can be noticed in the R-band polarimetric results. The substructures do not appear in the \textit{Planck} measurements as \textit{Planck} traces the large-scale component of the magnetic field. Also, due to the relatively poor resolution, the \textit{Planck} data could not reveal any effect on the magnetic field by the local kinematics and overall substructures at high-density regimes. Even the \textit{Planck} polarization vectors from outside of the 9$\times10^{20}$ cm$^{-2}$ contour that represent the \bpos in ICM (\ticm=190$\degr$) are correlated with both \tr and \tp implying that the cloud \bpos is threaded by the ICM \bpos surrounding the cloud.   

In Fig. \ref{fig:p_pa_sec_d} (a) we show the \bpos geometry of the head region of L1172/1174 inferred from our R-band and the \textit{Planck} polarization results. In Fig. \ref{fig:p_pa_sec_d} (b) and (c) we show the \tr and \pr values of the stars as a function of their angular separation from HD 200775, respectively. The \textit{Planck} polarimetric values (\tp and \pp\!\!) are also shown in the same figures. The primary goal here is to investigate whether the presence of HD 200775 has any effect on the magnetic field geometry around it. Two components of \bpos (\tr$\geqslant196\degr$ and $<196\degr$) are apparent in Fig. \ref{fig:p_pa_sec_d} (b) for the sources lying toward the head region. The changes in the \tr and \tp found beyond $\sim1\degr$ toward the tail are due to the curved geometry of \bpos, which is found to correlate well with the geometry of the cloud structure there. The median values of \tr lying toward the head region and having \tr$\geqslant196\degr$ and $<196\degr$ are found to be 211$\degr$ and 179$\degr$ respectively with the MAD of 8$\degr$ for both the distributions. The distribution of \tr in Fig. \ref{fig:p_pa_sec_d} (b) actually reflects the distribution of the magnetic field vectors on the cloud as shown in Fig. \ref{fig:p_pa_sec_d} (a) with 179$\degr$ component lying predominantly to the northwestern parts of HD 200775 and the 211$\degr$ component lying to the southeastern parts. Similar to Fig. \ref{fig:p_pa_sec_d} (b), a clear bimodal angle distribution (\tr$\geqslant196\degr$ and $<196\degr$) of the histogram of \tr in head region can be noticed in Fig. \ref{fig:cepheus_pol_YSO} (c). As evident from that figure, the histogram of \tr in the tail peaks at the median \tr (196$\degr$), almost overlapping with the higher \tr values ($\geqslant196\degr$), which is noticeable in Fig. \ref{fig:p_pa_sec_d} (b) also. Fig. \ref{fig:p_pa_sec_d} (c) is similar to Fig. \ref{fig:cepheus_pol_YSO} (b), which shows that \pp values are similar to the \pr values. While we obtained the bimodal distribution both in \tp and \tr, there is no such bimodality present in \pp and \pr\!\!.

The four observed YSO candidates in Table \ref{tab:yso_pol_results_table} are all found to be located toward the head region. As evident from Fig. \ref{fig:cepheus_pol_YSO} (b) and (c), the polarimetric results of all the YSO candidates are found to be consistent with other observed stars. The \pr and \tr of YSO candidates agree well with the median values of the same in the head region of Table \ref{tab:pol_results_table}, except the \tr of star ID 3. However, the latter falls well within the range of \tr of the head region. Except star ID 18 in Table \ref{tab:yso_pol_results_table}, \tr of other YSO candidates fall on the group of lower \tr ($<196\degr$) values.

To investigate the variation in the projected magnetic field orientations in more detail, we divided the head region into 8 equal sectors drawn within the circular area of 0.4$\degr$ radius centered at HD 200775 as shown in Fig. \ref{fig:p_pa_sec_d} (a). The division of the region is made with respect to the symmetry axis of the outflow cavity believed to have been carved out by HD 200775 \citep{1998A&A...339..575F}. The star formation process is ongoing mainly in this circular region. The median values of \pr (column 2), \tr (column 3) and the MAD (column 4) of the \tr for each sector are given in Table \ref{tab:mag_str}. The component showing the median value of 211$\degr$ is dominant toward the sectors 2, 3, and 4, which lie to the south and southeast of HD 200775. The median value of \pr for this component is 2.4\%. The component showing the median value of 179$\degr$ is largely distributed toward sectors 7 and 8, which lie to the north of HD 200775. The median value of \pr for this component is 1.6\%, which is relatively low ($\sim2\sigma$). Sectors 1, 5, and 6 show the presence of both the components. The median value of the \tr in these three regions is found to be $\sim200\degr$, which is roughly the average of the two (179$\degr$ and 211$\degr$) components. The deviation in the \tr ($\Delta\theta_\mathrm{R}$) is found to be the lowest toward the tail region and highest toward sectors 5 and 6.

\begin{figure}
	\centering
	\includegraphics[width=6cm, height=12cm]{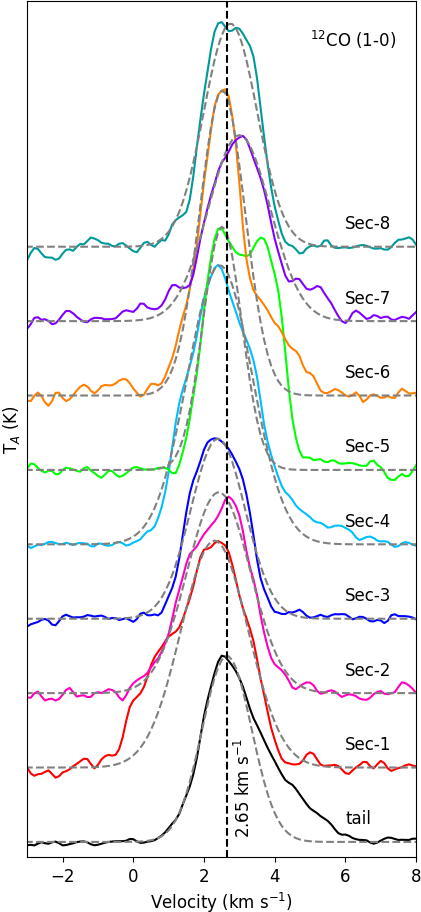}
	\caption{Average $^{12}$CO (J=1-0) line profiles for the eight sectors made toward the head. The average $^{12}$CO (J=1-0) line profile for the tail region is shown in black. The V$_{lsr}=$2.65 km~s$^{-1}$ is shown using a dashed vertical line. The dashed curves are the Gaussian fitted lines used to obtain the $\Delta$V.} \label{fig:vel_avg}
\end{figure}

In Fig. \ref{fig:vel_avg} we show the average $^{12}$CO (J=1-0) line profiles for the eight sectors identified toward the head region. The $^{12}$CO lines toward the lines-of-sight of the stars for which we have R-band polarization measurements are used for getting the average profiles. The average $^{12}$CO line profile for the tail region is also shown in Fig. \ref{fig:vel_avg}. The measurements having S/N $\geqslant$3 are used for generating the average profiles. The estimated V$_{lsr}$ (=2.65 km~s$^{-1}$) of the cloud complex is also identified with a vertical line to show the presence of redshfited and blueshfited velocity components in different sectors. The $^{12}$CO line toward the tail region is found to be peaking at the V$_{lsr}$ velocity with a FWHM ($\Delta$V) value of 1.7 km~s$^{-1}$ obtained from a Gaussian fit to the profile. However, the profile shows a line profile skewed to the redder velocities, which is more likely to be due to the presence of high-velocity (both blue- and red-shifted) gas.

The average $^{12}$CO line profiles for sectors 3, 4, and 7 are found to be consistent with a Gaussian shape, though presence of high-velocity components are seen toward most of the sectors similar to what we observed toward the tail region. While the line peaks toward sectors 1, 3, 4, and 5 are shifted toward the bluer velocities, the line toward sector 7 shows a shift toward the redder velocity with respect to the V$_{lsr}$. Sectors 2 and 6 are found to peak at the V$_{lsr}$ velocity. The line profiles at sectors 1 and 2 show additional components at bluer velocities, whereas a relatively narrow linewidth with high-velocity wings to the redder velocities are seen for the profile in sector 6. Compared to other sectors, sector 5 shows distinct profile with an additional velocity component to the redder side of the line center having both the peaks with comparable intensity. Thus, for sector 5, we fitted the observed line profile with the two Gaussian components and the linewidth ($\Delta$V$\approx$1.3 km~s$^{-1}$) corresponding to the component closest to the V$_{lsr}$ is used to estimate the magnetic field strength.

It is observed that the high-velocity gas is not widespread but localized. It is possible that the high-velocity gas present toward, especially, sectors 5 and 6 may be responsible for the disturbance of grain alignment and hence the relatively high $\Delta$\tr seen in these two regions where two components of \tr ($<196\degr$ and $\geqslant196\degr$) are present. The symmetry axis of the cavity, believed to be carved out by the outflow from HD 200775 \citep{1998A&A...339..575F}, is found to be in a direction almost perpendicular to the magnetic field direction. The cavity located to the northwest of HD 200775 is found to be more extended suggesting that this part is relatively more affected by the star than the southeastern cavity. Presence of high-velocity gas in sector 6 supports this observation. In sector 5 also, the presence of additional velocity components disturbs the surrounding material, which results in a higher dispersion in \tr. The additional velocity component is considered to originate from a loop structure close to L1172 (will be discussed in detail in Sharma et al. 2021, under preparation). The $\Delta$V values estimated for the lines in all sectors are listed in Table \ref{tab:mag_str}.

From Fig. \ref{fig:vel_avg}, it is evident that a majority of the average $^{12}$CO (J=1-0) line profiles seem to have multiple velocity components, though we selected the outer envelope of the cloud complex (where we carried out the R-band polarimetric study), which is of relatively low density. This arises a possibility of overestimation of $^{12}$CO linewidth. It would have been better to use $^{13}$CO, which was not available in our study. Though we had C$^{18}$O data toward this region, but being a high-density gas tracer, we could get emission only in the inner region of the cloud complex, where polarimetric measurements are not available. Therefore, we restrained the linewidth measurements using $^{12}$CO only.
\begin{table}  
    \begin{center}
    \footnotesize
        \caption{P\%, $\Delta\theta$, $\Delta$V, n$_{\text{H}_2}$ and B$_\mathrm{POS}$ in different sectors of L1172/1174 cloud complex.}\label{tab:mag_str}
        \begin{tabular}{ccccccc}\hline
        Id$^{\dagger}$ &$\mathrm{P_R}$&$\theta_{R}^{*}$&$\Delta\theta_{R}$&$\Delta \text{V}$&n$_{\text{H}_2}$& B$_{\text{POS}}$\\
         & (\%) & ($\degr$) & ($\degr$) & (kms$^{-1}$) & (cm$^{-3}$) & ($\mu$G)\\
         (1)&(2)&(3)&(4)&(5)&(6)&(7)\\\hline
        1 & 2.6$\pm$0.7& 198$\pm$8 & 6$\pm$2 &2.2$\pm$0.2 &93$\pm$38 &33$\pm$13\\
        2 & 2.2$\pm$0.8& 213$\pm$7 & 6$\pm$2 &2.1$\pm$0.1 &91$\pm$37 &31$\pm$12\\
        3 & 2.2$\pm$0.8& 212$\pm$7 & 5$\pm$2 &1.8$\pm$0.1 &90$\pm$36 &32$\pm$14\\
        4 & 2.7$\pm$0.6& 219$\pm$8 & 7$\pm$1 &2.2$\pm$0.2 &101$\pm$40 &29$\pm$8\\
        5 & 3.3$\pm$0.8& 200$\pm$11 & 10$\pm$1 &1.3$\pm$0.2 &121$\pm$49 &13$\pm$4\\
        6 & 2.0$\pm$0.4& 200$\pm$13 & 12$\pm$1 &1.5$\pm$0.1 &70$\pm$28 &10$\pm$2\\
        7 & 1.4$\pm$0.3& 174$\pm$8 & 6$\pm$2 &2.1$\pm$0.1 &67$\pm$27 &27$\pm$10\\
        8 & 1.3$\pm$0.5& 178$\pm$8 & 6$\pm$2 &1.8$\pm$0.1 &93$\pm$37 &27$\pm$11\\ 
        tail & 2.0$\pm$0.4& 196$\pm$6 & 4$\pm$1 &1.7$\pm$0.2 & 113$\pm$45 &42$\pm$14\\\hline
        \end{tabular}\\
    \end{center}
        $^{\dagger}$ The identification numbers of sectors 1-8 as shown in Fig. \ref{fig:p_pa_sec_d} (a).\\ 
        $^{*}$ Optical R-band measurements.
\end{table}

\subsection{Magnetic field strength in L1172/1174}

The strength of the plane-of-sky component of the magnetic field was estimated using the modified Chandrasekhar-Fermi (CF) relation \citep{1953ApJ...118..113C,2001ApJ...546..980O},
\begin{equation}
\mathrm{B}_\mathrm{POS} = 9.3 \left[ \frac{\mathrm{n}_\mathrm{H_{2}}}{\mathrm{cm}^{-3}} \right]^{1/2} \left[ \frac{\Delta \mathrm{V}}{\mathrm{km s}^{-1}} \right] \left[\frac{\Delta \theta}{1^{\circ}} \right]^{-1}~~\mu G, \label{equ:B_field}
\end{equation}
where n$_\mathrm{H_{2}}$ represents the volume density of hydrogen gas in molecular clouds, $\Delta$V is the FWHM obtained from the velocity dispersion $\sigma$V (\!$\sqrt{8ln2}$ $\sigma$V), and $\Delta\theta$ is the dispersion in the position angles. The CF method is applicable for $\Delta\theta<\!25^{\circ}$. This method suggests that the strength can be estimated by the analysis of small-scale randomness in the magnetic field lines. The line-of-sight velocity dispersion causes an irregular scatter in the polarization position angles under the assumptions that there is a mean field component in the area of interest, that the turbulence responsible for the magnetic field perturbations is isotropic, and that there is equipartition between the turbulent kinetic and magnetic energy \citep{2001ApJ...561..800H}. The CF method of estimating B$_\mathrm{POS}$ is a statistical method that may be in error by a factor of $\sim$2 for individual clouds.

The observed $\Delta\theta_{R_{obs}}$ is the joint contribution of both the intrinsic dispersion ($\Delta\theta_{R}$) and measurement uncertainties ($\sigma_{\theta_{R}}$) \citep{2001ApJ...561..864L}. Therefore, we obtained the $\Delta\theta_{R}$ using $\Delta\theta_{R}$ = $\sqrt{(\Delta\theta_{R_{obs}}^{2}-\sigma_{\theta_{R}}^{2}}$). The $\sigma_{\theta_{R}}$ is the average uncertainty, which was estimated using $\sigma_{\theta_{R}}$ = $\sum\sigma_{\theta_{R_{i}}}$/N, where  $\sigma_{\theta_{R_{i}}}$ is the measured uncertainty of the i$^{th}$ star’s polarization angle and N is the total number of stars. The values of $\Delta$V for sectors 1-8 and for the tail, as identified in Fig. \ref{fig:p_pa_sec_d} (a), are obtained by fitting Gaussian profiles (single component) to the average $^{12}$CO spectrum generated for the individual regions. The estimated values of $\Delta$V are given in the column 5 of Table \ref{tab:mag_str}. $\Delta$V is the combination of both thermal ($\Delta$V$_{th}$) and nonthermal ($\Delta$V$_{NT}$) gas components along the line-of-sight ($\Delta$V=$\sqrt{\Delta\mathrm{V}_{th}^{2}+\Delta \mathrm{V}_{NT}^{2}}$) \citep{1983ApJ...270..105M}. $\Delta$V$_{th}$ can be estimated from $\sigma_{th}=\sqrt{kT/m}$, where $\sigma_{th}$ is the thermal velocity dispersion of the gas, k is the Boltzmann constant, T is the kinetic temperature and m is molecular weight of the gas. T = 10 K was assumed to estimate the $\Delta$V$_{th}$ in our analysis. We removed the thermal component from the observed linewidth in each sector. The angular extent of the head within the 9$\times10^{20}$ cm$^{-2}$ contour is $\sim0.6\degr$, which is $\sim$ 3.5 pc considering its distance as $\sim$ 335 pc \citep{2020MNRAS.494.5851S}. For the tail, based on the same contour level, the angular extent was found to be $0.5\degr$, which translates to $\sim$ 2.9 pc at 335 pc. Assuming a similar line-of-sight extent for the cloud and using the hydrogen column density values obtained toward individual stars for which we made R-band polarization measurements, we calculated an average value for n$_\mathrm{H_{2}}$ for each sectors and the tail region. The n$_\mathrm{H_{2}}$ thus obtained are given in column 6 of Table \ref{tab:mag_str}. The uncertainty in the measurement of n$_\mathrm{H_{2}}$ is taken as 40\% of the value \citep{2016MNRAS.461...22P}. The \bpos strength calculated for the head and the tail regions are given in column 7 of Table \ref{tab:mag_str}. The magnetic field strength is found to be weakest in sector 6, which is due to the combined effect of narrow $\Delta$V and relatively high value of $\Delta\theta_{R}$. The average value of magnetic field strength for the entire cloud is found to be $\sim$30 $\mu$G. 

\subsection{Large-scale magnetic field and bulk motion}

\begin{figure*}
    \includegraphics[width=9.3cm, height=8.3cm]{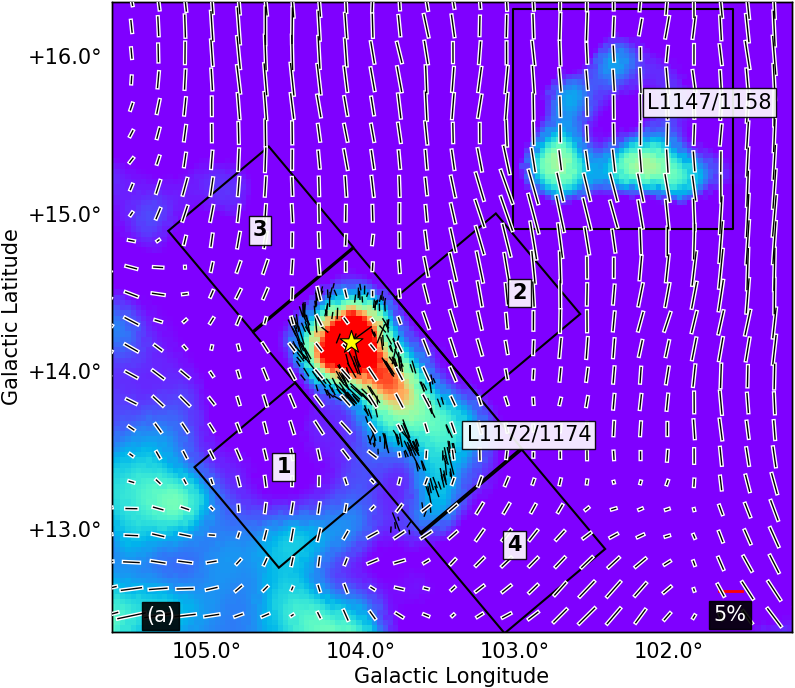}
    \hspace{0.5cm}
    \includegraphics[width=7.6cm, height=8.3cm]{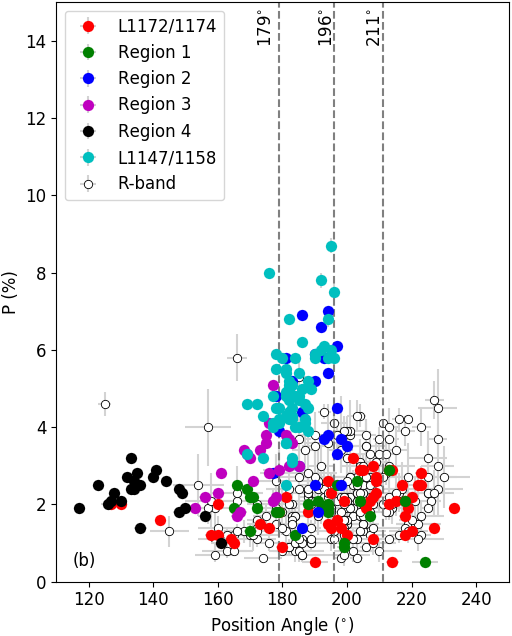}
    \caption{Variation in the projected magnetic field geometry toward L1172/1174 shown in a wider (4$\degr\times4\degr$) area. (\textbf{a}) Magnetic field map (lines in black with white border) of the region covering L1147/1158 and L1172/1174 inferred from the \textit{Planck} polarization measurements. The magnetic field map of L1172/1174 produced using our R-band polarization measurements are also shown using black lines. (\textbf{b}) Regions from which we obtained \textit{Planck} polarization measurements shown as boxes (1, 2, 3, 4, L1172/1174, and L1147/1158), plotted along with our R-band polarization measurements in L1172/1174.} \label{fig:pol_zones}
\end{figure*}

Overall, the \bpos geometry in L1172/1174 cloud complex inferred from both the R band and \textit{Planck} displays an ``S" shape morphology with the field lines changing their orientations smoothly as we move from the head to the tail region with \bpos oriented almost parallel to the main filament of the cloud as traced by the 15$\times10^{20}$ cm$^{-2}$ contour as seen in Fig. \ref{fig:cepheus_pol_YSO} (a). In Fig. \ref{fig:pol_zones} (a) we show the magnetic field geometry of the region surrounding L1172/1174 and L1147/1158 ($1.4\degr\times1.4\degr$) cloud complexes. The large-scale survey carried out in $^{13}$CO (J=1-0) line of the Cepheus and Cassiopia regions by \citet{1997ApJS..110...21Y} suggests that both the complexes share similar radial velocities ($2.7-2.9$ km~s$^{-1}$) implying that the two regions are both spatially and kinematically connected. The distance of 340$\pm$3 pc to L1147/1158 estimated using the \textit{Gaia} DR2 parallax and the proper motion measurements of the YSOs associated with the complex \citep{2020A&A...639A.133S} suggests that both L1147/1158 and L1172/1174 are roughly at similar distance from us. At this distance, the spatial separation between the two complexes is $\sim9$ pc. 

In Fig. \ref{fig:pol_zones} (b) we plot \pp versus \tp from four regions (square boxes of size $0.8\degr\times0.8\degr$) selected around L1172/1174 cloud complex as shown in Fig. \ref{fig:pol_zones} (a) to examine the large-scale magnetic field orientation in the surrounding medium and how it is related to the cloud field lines. The broken lines identify the median values of the two position angle components identified from \tr toward the head and the median value of the \tr toward the tail as shown in Fig. \ref{fig:p_pa_sec_d} (b). The median and MAD values of \tp toward L1148/1157 are 182$\degr$ and 4$\degr$, respectively. This is comparable to the 179$\degr$ component of \tr in the northwestern part of the head region of L1172/1174. The \tp distribution found toward the regions 1, 2, and 3 resembles the \tp distribution found over the whole cloud complex L1172/1174. This is true for both the R band and the \textit{Planck} polarization results suggesting that the cloud is permeated with the ICM magnetic field. The R-band polarization seems to be consistent with the ICM large-scale magnetic field (derived from \textit{Planck}) but shows substructures as it follows closely the column density contours (see Fig. \ref{fig:cepheus_pol_YSO} (a)). The median values of \tp for regions 1, 2, 3, and 4 are 193$\degr\pm$14, 191$\degr\pm$5, 175$\degr\pm$6, and 135$\degr\pm$7, respectively. The \tp in regions 1 and 2 are consistent with the median value of \tr obtained for the tail region. The median value of 175$\degr$ obtained for the region 3 is consistent with the 179$\degr$ component obtained toward sectors 7 and 8 in the head region suggesting that the magnetic field in the head inferred from the R-band polarization measurements is smoothly joining with the ICM field lines inferred from the \textit{Planck} polarization measurements. The 135$\degr$ component obtained for the region 4 resembles the change in the magnetic field geometry seen toward the tail part in R-band polarization measurements again suggesting a smooth merger between the cloud and ICM field lines. The \tp is found to be relatively scattered in region 1 and in its northeastern part, the alignment of \tp is similar to \tr of sectors 2 and 3. Globally, regions in the northwest (L1147/1158, L1172/1174, 2, and 3) show \tp$\sim180\degr$ whereas in the southeastern part (including 1 and 4) \tp is highly disturbed.   

Based on the Histogram of Relative Orientations analysis \citep[HRO; ][]{2013ApJ...774..128S} with decaying supersonic turbulence, it was shown that in the high-magnetization scenario, the magnetic field changes direction from being parallel to perpendicular with respect to the density structures. The relative orientation becomes random or parallel in the intermediate- or low-magnetization scenario suggesting that the strength of magnetic field may also play a crucial role. The relative orientation is also found to change progressively with increasing column density, from mostly parallel or having no preferred orientation to mostly perpendicular \citep{2016A&A...586A.138P}. Applying HRO analysis, \citet{2019A&A...629A..96S} presented the orientation of \bpos for the Cepheus flare region using the \textit{Planck} polarization data \citep[Fig. 3: ][]{2019A&A...629A..96S}. It was noted that the elongated structure of the cloud complex and \bpos are almost parallel across all the column density values in L1172/1174 cloud complex while they are perpendicular in the L1147/1158 cloud complex, which is clearly visible in Fig. \ref{fig:pol_zones} (a). There is also a marked difference between the magnetic field orientation to the northwestern and southeastern parts of L1172/1174. While the magnetic fields show an almost uni-directional pattern (mostly along 180$\degr$) in the northwestern parts of L1172/1174 where L1147/1158 is also located, to the southeastern parts, the magnetic field lines show more twists and turns.

Studies conducted to investigate the relative orientation between the filamentary structure and the magnetic field in a sample of molecular clouds associated with the Gould Belt report a bimodal distribution of the angles between filaments and magnetic fields with the offsets being either parallel or perpendicular \citep{2013MNRAS.436.3707L, 2019ApJ...871L..15G}. Results from simulations \citep{2015MNRAS.452.2410S} and observational studies \citep{2017NatAs...1E.158L} suggest that the bimodality may have implications in cloud's evolution and subsequent star formation process. The molecular clouds with the long axes perpendicular to the magnetic field directions are shown to have more evenly distributed linear mass across the field lines \citep{2019MNRAS.484.3604L}. Also, these clouds consistently show a lower star formation rate (SFR)/mass for clouds \citep{2017NatAs...1E.158L}. The perpendicular alignment of magnetic field lines possesses a significantly higher flux compared to the parallel orientation and thus provides a stronger support to the molecular clouds against self-gravity \citep{2017NatAs...1E.158L}. A total of twelve molecular clouds with their SFR/mass and relative orientations of their long axes with magnetic field lines are listed by \cite{2017NatAs...1E.158L}, based on optical and \textit{Planck} measurements. Of these twelve clouds, six have projected magnetic field aligned perpendicular to the longer axes, while this alignment is parallel for other six clouds. \cite{2017NatAs...1E.158L} found significant difference in SFR/cloud mass in these two sets of orientations (see Table \ref{tab:cloud_b}). Our study makes further addition to the findings by \cite{2017NatAs...1E.158L}. 

\begin{table}  
    \begin{center}
    \small
        \caption{SFR per unit cloud mass values of molecular clouds from \cite{2017NatAs...1E.158L} and from our study.}\label{tab:cloud_b}
        \begin{tabular}{lccc}\hline
        Cloud & SFR per unit& SFR per unit& SFR per unit\\
        name & cloud mass$^{\dagger}$& cloud mass$^{\ddagger}$& cloud mass$^{*}$\\\hline
        \multicolumn{4}{c}{Perpendicular alignment}\\
        IC 5146&-&0.380$\pm$0.18&-\\
        Pipe Nebula&2.81&-&-\\ 
        Orion&4.17&-&-\\
        Chamaeleon&-&1.03$\pm$0.48&-\\
        Taurus&4.76&0.140&-\\
        Lupus I&4.00&0.630$\pm$0.52&-\\
        L1147/1158$^{*}$&-&-&0.4$\pm$0.3\\\hline
        \multicolumn{4}{c}{Parallel alignment}\\
        Lupus II-VI&6.97&1.85$\pm$0.74&-\\
        Corona Australis&9.69&3.66$\pm$2.4&-\\
        Cepheus&-&1.13$\pm$0.62&-\\
        Ophiuchus&6.10&2.32$\pm$1.8&-\\
        Aquila&-&1.48$\pm$0.8&-\\
        Perseus&7.98&1.46$\pm$1.1&-\\
        L1172/1174$^{*}$&-&-&2.0$\pm$1.3\\\hline
        \end{tabular}
    \end{center}
$^{\dagger}$ Obtained from \cite{2010ApJ...724..687L}.\\
$^{\ddagger}$ Obtained from \cite{2010ApJ...723.1019H}.\\
$^{*}$ Obtained from this work.
\end{table}
	
We estimated the SFR in L1172/1174 complex using the total number of YSOs identified toward the cloud and by assuming a mean YSO mass of 0.5$\pm$0.1 M$_{\odot}$ and age of 2$\pm$1 Myr \citep{2009ApJS..181..321E, 2010ApJ...723.1019H, 2014ApJ...782..114E, 2017NatAs...1E.158L}. A total of 74 YSOs have been identified so far toward L1172/1174 complex \citep{2020MNRAS.494.5851S}. Based on their positions on the color-magnitude diagram produced using the data from the \textit{Gaia} DR2, \cite{2020MNRAS.494.5851S} found an age of $\sim$1-2 Myr for the YSOs, which is consistent with the median age of $\sim$1.6 Myr determined by \citet{2009ApJS..185..451K}, and thus, in turn, similar to the age we assumed for the estimation of SFR in this cloud complex. Also, the majority of the sources identified so far in L1172/1174 complex are of low-mass M types (0.1-0.7 M$_{\odot}$), which is found to be consistent with our assumptions. The SFR thus calculated is found to be 19$\pm$10 M$_{\odot}$ Myr$^{-1}$. The mass of the cloud was estimated by summing up all those pixels having the hydrogen column density threshold of 9$\times10^{20}$ cm$^{-2}$ (A$_{V}\gtrsim$1.0 magnitude) and using it in the expression M $=$ N(H$_{2}$)m$_{H}\mu$A \citep{2018A&A...615A.125B}, where M is the cloud mass per pixel, N(H$_{2}$) indicates the H$_{2}$ column density, m$_{H}$ represents mass of a hydrogen atom, $\mu=2.86$, the mean particle mass (assuming $\sim$70\% H$_{2}$ by mass), and A represents the area of each pixel. We estimated the area A using ($\pi$/180/3600)$^{2}$ D(cm)$^{2}$ R($^{\prime\prime}$)$^{2}$, where D is the distance to the cloud complex and R($^{\prime\prime}$) is the size of one pixel. We estimated a cloud mass of 950$\pm$380 M$_{\odot}$. The error is estimated by propagating the uncertainties in the column density estimation and in the distance. Recently, \cite{2020ApJ...904..172D} estimated the mass of L1172/1174 as 1000 M$_{\odot}$ with A$_{V}>$1, which is consistent with the mass we obtained. Thus the SFR/mass estimated for L1172/1174 is 2.0$\pm$1.3 \%Myr$^{-1}$, which is comparable with the mean value 1.98$\pm$0.92 \%Myr$^{-1}$ obtained by \cite{2017NatAs...1E.158L}, for the clouds with parallel alignments (e.g., Ophiuchus, Corona Australis, Aquila, Perseus etc.). Similarly, we estimated the SFR/mass of the neighboring cloud complex L1147/1158, which has its long axis almost perpendicular to the projected magnetic field direction. Considering the same hydrogen column density threshold (9$\times10^{20}$ cm$^{-2}$) condition, the mass of L1147/1158 is estimated to be 800$\pm$320 M$_{\odot}$, which is consistent with the mass 790 M$_{\odot}$, obtained by \cite{2020ApJ...904..172D}. A total of 14 YSO candidates \citep{2009ApJS..185..198K} including PV Cep are distributed within the defined threshold. The SFR is computed as 3.5$\pm$1.9 M$_{\odot}$ Myr$^{-1}$. Therefore, the estimated SFR/mass for L1147/1158 is 0.4$\pm$0.3 \%Myr$^{-1}$, which is significantly lower than the same obtained for L1172/1174. But it is consistent with the mean value for the clouds having magnetic field lines perpendicular to their long axes \citep[0.620$\pm$0.37 \% Myr$^{-1}$;][]{2017NatAs...1E.158L}. These results further support the claim by \cite{2017NatAs...1E.158L} that if the magnetic field lines are aligned parallel to the cloud's long axis (like in L1172/1174, Ophiuchus, Aquila, Perseus etc.), the SFR is found to be higher than in the clouds where the magnetic field lines are aligned perpendicular to the cloud's long axis (like in L1147/1158, IC 5146, Pipe Nebula, Taurus etc.).

\section{Summary and conclusion} \label{sec:sum_con}
We present the results of our R-band polarization measurements of 249 sources projected on the cloud complex L1172/1174. Combining our results with those from the \textit{Planck} polarization measurements of the region containing the complex, we studied the magnetic field geometry of the cloud and its relationship with the ICM magnetic field. We summarize the results obtained from this work below:

\begin{itemize}
    \item The magnetic field geometry inferred from our R-band polarization measurements and from \textit{Planck} are found to be in good agreement throughout the cloud. The magnetic field is found to be smooth (not chaotic) and oriented along the hub-filament structure of the cloud. The only changes noticed are toward the extreme ends of the head and the tail regions where the magnetic field lines are found to join smoothly with the ambient magnetic fields in the ICM. 
    \item Overall, there is not much effect of the presence of HD 200775 on the magnetic field geometry of the surrounding region except toward the northwestern part of HD 200775 where field lines are showing a relatively large dispersion in the magnetic field vectors. The $^{12}$CO line profile shows a presence of high-velocity clouds in this region, which may be responsible for disturbing the magnetic field geometry.
    \item The mean magnetic field strength for the entire cloud was found to be $\sim$30 $\mu$G. Globally, the magnetic field geometry inferred from the \textit{Planck} polarization is found to be oriented along a mean position angle of $180^{\degr}$ toward the northwestern side of L1172/1174 where the cloud complex L1147/1158 is located. At the location of the cloud, the magnetic field lines change to a mean position angle of $200^{\degr}$ and then to the southeastern side of L1172/1174, the magnetic field lines show more twists and turns.
    \item The higher SFR/mass of 2.0$\pm$1.3 \%Myr$^{-1}$ found for L1172/1174, compared to 0.4$\pm$0.3 \%Myr$^{-1}$ for L1147/1158, is consistent with the earlier results, which suggests that the molecular clouds with magnetic field lines oriented parallel to the cloud elongation are found to show relatively high values of SFR compared to those with field lines perpendicular to the cloud elongation. 
\end{itemize}
\noindent
As a whole, this study helps us to understand the morphology of the projected magnetic field geometry toward the inner high-density as well as the outer low-density regions of the L1172/1174 cloud complex and also its impact in the SFR. 


\begin{acknowledgements}

We thank Dr. Archana Soam and Neha Sharma, and all the supporting staff at ARIES, Nainital who made these observations possible. C.W.L was supported by the Basic Science Research Program (2019R1A2C1010851) through the National Research Foundation of Korea. TG acknowledges support from the Science and Engineering Research Board of the Department of Science and Technology, Govt. of India, grant number SERB/ECR/2018/000826. This work has made use of data from the following sources: (1) the Planck Legacy Archive (PLA) contains all public products originating from the Planck mission, and we take the opportunity to thank ESA/Planck and the Planck collaboration for the same. (2) European Space Agency (ESA) mission {\it Gaia} (\url{https://www.cosmos.esa.int/gaia}), processed by the {\it Gaia} Data Processing and Analysis Consortium (DPAC, \url{https://www.cosmos.esa.int/web/gaia/dpac/consortium}). Funding for the DPAC has been provided by national institutions, in particular the institutions participating in the {\it Gaia} Multilateral Agreement. This research also has made use of the SIMBAD database, operated at CDS, Strasbourg, France.

\end{acknowledgements}

\bibliographystyle{aa} 
\bibliography{reference} 

\begin{appendix}
\section{Tables}
\begin{table*}
\caption{R-band polarimetric results of 249 stars observed toward L1172/1174.}\label{tab:res249}
\footnotesize
\begin{tabular}{cc} 
\renewcommand{\arraystretch}{1.3}
\begin{tabular}[h]{cccccr}  \hline
Star & $l$ & $b$  & P$\pm$$\sigma_{P}$ & $\theta$$\pm$$\sigma_{\theta}$ & $d$$\pm$$\sigma_{d}$\\ 
Id  &   ($\degr$) &    ($\degr$)&  (\%)    & ($\degr$) & (pc)\\\hline
1&103.369563&13.535764&1.8$\pm$0.4&191$\pm$5& 1335$_{-64}^{+71}$\\ 
2&103.376949&13.534869&1.9$\pm$0.4&192$\pm$4&  461$_{-4}^{+4}$\\ 
3&103.388335&13.423807&3.6$\pm$1.0&194$\pm$7& 1269$_{-70}^{+78}$\\ 
4&103.391221&13.371439&2.0$\pm$0.1&189$\pm$1& 1295$_{-40}^{+43}$\\ 
5&103.396714&13.370003&0.8$\pm$0.3&184$\pm$6&  882$_{-18}^{+19}$\\ 
6&103.399226&13.548511&1.8$\pm$0.5&188$\pm$6& 2088$_{-139}^{+160}$\\ 
7&103.403268&13.643440&2.8$\pm$0.5&205$\pm$5& 2619$_{-222}^{+266}$\\ 
8&103.405470&13.542955&2.3$\pm$0.4&188$\pm$4& 5459$_{-782}^{+1031}$\\ 
9&103.407969&13.506287&2.1$\pm$0.1&193$\pm$1&   475$_{-4}^{+4}$\\ 
10&103.413906&13.654661&1.5$\pm$0.5&200$\pm$7& 6033$_{-1111}^{+1509}$\\ 
11&103.420810&13.460331&1.9$\pm$0.4&200$\pm$5&  469$_{-4}^{+4}$\\ 
12&103.425854&13.408299&2.7$\pm$0.9&185$\pm$8& 3717$_{-355}^{+434}$\\ 
13&103.427039&13.500412&2.4$\pm$0.1&191$\pm$1&  659$_{-6}^{+6}$\\ 
14&103.432185&13.336416&2.1$\pm$0.1&197$\pm$1&  457$_{-5}^{+5}$\\ 
15&103.432276&13.357525&2.1$\pm$0.1&191$\pm$1&  552$_{-8}^{+8}$\\ 
16&103.433914&13.462983&1.7$\pm$0.1&202$\pm$1&  665$_{-10}^{+10}$\\ 
17&103.445397&13.488413&1.8$\pm$0.6&180$\pm$7& 2957$_{-179}^{+203}$\\ 
18&103.446267&13.324590&2.4$\pm$0.5&193$\pm$5& 3176$_{-198}^{+226}$\\ 
19&103.446568&13.834850&1.3$\pm$0.4&196$\pm$6& 3428$_{-271}^{+319}$\\ 
20&103.453856&13.366588&2.5$\pm$0.6&191$\pm$6& 2097$_{-196}^{+240}$\\ 
21&103.466449&13.297720&1.5$\pm$0.3&200$\pm$4&  779$_{-20}^{+21}$\\ 
22&103.469665&13.821971&1.6$\pm$0.1&200$\pm$2&  457$_{-3}^{+3}$\\ 
23&103.484429&13.259809&2.2$\pm$0.5&200$\pm$5&  658$_{-13}^{+13}$\\ 
24&103.496048&13.316274&2.2$\pm$0.3&201$\pm$3& 1892$_{-175}^{+213}$\\ 
25&103.500559&13.963767&0.8$\pm$0.3&202$\pm$6& 1800$_{-52}^{+55}$\\ 
26&103.513589&13.762300&2.2$\pm$0.4&207$\pm$5&  765$_{-14}^{+14}$\\ 
27&103.515728&13.971093&1.8$\pm$0.2&191$\pm$2& 6644$_{-753}^{+939}$\\ 
28&103.517554&13.306443&2.3$\pm$0.5&195$\pm$5& 3523$_{-330}^{+402}$\\ 
29&103.522295&13.800443&2.0$\pm$0.2&192$\pm$2& 1588$_{-80}^{+89}$\\ 
30&103.524179&13.922084&2.2$\pm$0.3&188$\pm$3&  804$_{-24}^{+26}$\\ 
31&103.527084&13.154722&1.7$\pm$0.2&184$\pm$3&  879$_{-12}^{+13}$\\ 
32&103.528496&13.281828&1.9$\pm$0.2&192$\pm$3& 1014$_{-26}^{+27}$\\ 
33&103.530729&13.817294&2.2$\pm$0.2&199$\pm$2&  426$_{-3}^{+3}$\\ 
34&103.531432&13.964377&2.1$\pm$0.1&189$\pm$1& 2415$_{-196}^{+232}$\\ 
35&103.532947&13.236985&3.5$\pm$0.5&190$\pm$4& 2165$_{-204}^{+250}$\\ 
36&103.536011&13.799642&3.1$\pm$0.2&204$\pm$1& 2322$_{-142}^{+161}$\\ 
37&103.547924&13.271056&2.2$\pm$0.1&198$\pm$1& 1146$_{-32}^{+34}$\\ 
38&103.561086&13.923598&1.3$\pm$0.3&207$\pm$5&  773$_{-12}^{+13}$\\ 
39&103.561332&13.119313&1.1$\pm$0.3&159$\pm$5& 1108$_{-43}^{+47}$\\ 
40&103.569513&13.423153&1.9$\pm$0.1&202$\pm$1&  751$_{-7}^{+7}$\\ 
41&103.576086&13.925862&1.8$\pm$0.4&195$\pm$5& 1175$_{-30}^{+32}$\\ 
42&103.577631&13.424342&1.9$\pm$0.6&195$\pm$6& 4283$_{-686}^{+943}$\\ 
43&103.578790&13.925869&2.0$\pm$0.4&205$\pm$4&  630$_{-140}^{+247}$\\ 
44&103.579566&13.171484&1.0$\pm$0.2&189$\pm$5&  428$_{-5}^{+5}$\\ 
45&103.610325&13.569492&1.4$\pm$0.3&199$\pm$4& 2222$_{-129}^{+145}$\\ 
46&103.613454&13.416918&1.7$\pm$0.5&191$\pm$7& 6817$_{-1261}^{+1680}$\\ 
47&103.615140&13.553508&2.2$\pm$0.4&193$\pm$4&  347$_{-3}^{+3}$\\ 
48&103.615277&13.441983&2.7$\pm$0.8&207$\pm$7& 2376$_{-270}^{+345}$\\ 
49&103.620378&13.534314&2.5$\pm$0.4&202$\pm$4&  889$_{-14}^{+14}$\\ 
50&103.622884&13.548024&2.1$\pm$0.4&203$\pm$5& 1255$_{-35}^{+37}$\\ 
51&103.625378&13.540711&2.5$\pm$0.6&205$\pm$6&  700$_{-11}^{+11}$\\ 
\hline
\end{tabular}
\begin{tabular}{cccccr}  \hline
Star & $l$ & $b$  & P$\pm$$\sigma_{P}$ & $\theta$$\pm$$\sigma_{\theta}$ & $d$$\pm$$\sigma_{d}$\\ 
Id  &   ($\degr$) &    ($\degr$)&  (\%)    & ($\degr$) & (pc)\\\hline
52&103.631329&13.531912&3.4$\pm$0.6&215$\pm$4&  778$_{-27}^{+29}$\\ 
53&103.632864&14.049986&2.5$\pm$0.4&196$\pm$4& 1042$_{-26}^{+28}$\\ 
54&103.638274&13.523002&2.3$\pm$0.2&205$\pm$2&  498$_{-40}^{+48}$\\ 
55&103.648487&14.041594&1.5$\pm$0.3&198$\pm$5& 1429$_{-50}^{+54}$\\ 
56&103.653112&13.516389&2.3$\pm$0.5&203$\pm$5& 1841$_{-102}^{+114}$\\ 
57&103.657643&13.573459&2.7$\pm$0.4&201$\pm$4& 1874$_{-127}^{+146}$\\ 
58&103.662384&13.512556&1.6$\pm$0.2&199$\pm$3&  856$_{-11}^{+11}$\\ 
59&103.662619&13.545521&1.8$\pm$0.1&207$\pm$2&  355$_{-2}^{+2}$\\ 
60&103.664697&13.001426&0.7$\pm$0.2&186$\pm$5&  741$_{-10}^{+11}$\\ 
61&103.666569&13.051837&1.3$\pm$0.4&145$\pm$6& 2075$_{-152}^{+177}$\\ 
62&103.680989&13.501700&1.3$\pm$0.1&205$\pm$2& 1686$_{-98}^{+110}$\\ 
63&103.681499&13.550944&2.0$\pm$0.3&206$\pm$4& 1293$_{-38}^{+40}$\\ 
64&103.682929&13.658846&3.8$\pm$0.2&201$\pm$1& 1709$_{-76}^{+83}$\\ 
65&103.683858&13.653563&1.7$\pm$0.3&204$\pm$5& 6516$_{-1141}^{+1528}$\\ 
66&103.685928&13.652803&2.7$\pm$0.3&203$\pm$2& 1847$_{-79}^{+86}$\\ 
67&103.690315&13.629575&3.5$\pm$0.6&206$\pm$4& 1208$_{-31}^{+32}$\\ 
68&103.691753&13.496346&1.1$\pm$0.3&195$\pm$5& 1386$_{-46}^{+50}$\\ 
69&103.699510&13.625001&4.0$\pm$0.5&223$\pm$3&  644$_{-36}^{+40}$\\ 
70&103.712075&14.044119&1.4$\pm$0.2&189$\pm$3& 4309$_{-782}^{+1100}$\\ 
71&103.717844&14.098736&1.0$\pm$0.1&189$\pm$3&  785$_{-11}^{+11}$\\ 
72&103.718710&13.585106&2.6$\pm$0.8&191$\pm$7& 2087$_{-284}^{+382}$\\ 
73&103.727308&14.084367&3.7$\pm$0.6&185$\pm$4& 1888$_{-271}^{+371}$\\ 
74&103.728869&14.078656&2.3$\pm$0.5&188$\pm$6&  722$_{-20}^{+22}$\\ 
75&103.730889&14.107847&3.8$\pm$0.4&206$\pm$3& 1903$_{-641}^{+1205}$\\ 
76&103.735578&13.086863&0.6$\pm$0.1&198$\pm$2&  388$_{-4}^{+4}$\\ 
77&103.737300&13.592209&1.9$\pm$0.2&194$\pm$3&  946$_{-16}^{+17}$\\ 
78&103.740288&14.046160&3.8$\pm$1.0&198$\pm$7& 3209$_{-395}^{+511}$\\ 
79&103.743451&13.589382&2.2$\pm$0.6&194$\pm$6& 5048$_{-854}^{+1172}$\\ 
80&103.743481&14.053415&3.2$\pm$0.5&200$\pm$4& 2627$_{-468}^{+683}$\\ 
81&103.750202&13.621534&1.9$\pm$0.4&214$\pm$5& 1103$_{-45}^{+49}$\\ 
82&103.751125&13.066935&3.9$\pm$0.8&201$\pm$5& 1918$_{-193}^{+240}$\\ 
83&103.751810&14.198525&2.0$\pm$0.2&214$\pm$3&  539$_{-7}^{+7}$\\ 
84&103.753142&14.055722&3.9$\pm$0.4&200$\pm$3&  689$_{-17}^{+18}$\\ 
85&103.753376&14.158306&5.8$\pm$0.6&166$\pm$3&  670$_{-26}^{+29}$\\ 
86&103.757171&13.044135&1.1$\pm$0.1&186$\pm$2& 1510$_{-57}^{+62}$\\ 
87&103.759157&14.202982&1.9$\pm$0.6&195$\pm$8& 1140$_{-43}^{+47}$\\ 
88&103.771324&14.083528&2.3$\pm$0.5&216$\pm$6& 1787$_{-150}^{+179}$\\ 
89&103.779461&14.166562&1.3$\pm$0.4&199$\pm$8& 3176$_{-221}^{+255}$\\ 
90&103.784884&13.017830&1.0$\pm$0.2&176$\pm$3& 2308$_{-92}^{+99}$\\ 
91&103.786605&14.168643&0.9$\pm$0.1&206$\pm$2&  756$_{-15}^{+16}$\\ 
92&103.788275&14.044542&3.0$\pm$0.8&213$\pm$8&  380$_{-7}^{+8}$\\ 
93&103.795807&14.224410&2.0$\pm$0.6&208$\pm$9&  915$_{-31}^{+33}$\\ 
94&103.795808&14.041406&3.7$\pm$0.3&210$\pm$2& 2489$_{-255}^{+316}$\\ 
95&103.799506&14.333130&1.1$\pm$0.2&172$\pm$6&  338$_{-4}^{+4}$\\ 
96&103.804584&14.310143&2.4$\pm$0.6&185$\pm$7& 1777$_{-161}^{+195}$\\ 
97&103.804980&14.054295&3.3$\pm$0.4&212$\pm$3& 3635$_{-463}^{+603}$\\ 
98&103.807211&14.052721&4.1$\pm$0.1&211$\pm$1& 1267$_{-44}^{+47}$\\ 
99&103.810584&14.308802&0.8$\pm$0.3&210$\pm$7&  734$_{-18}^{+19}$\\ 
100&103.814455&13.573526&2.7$\pm$0.9&190$\pm$9& 388$_{-8}^{+8}$\\ 
101&103.817286&13.567570&4.1$\pm$0.6&196$\pm$4&4632$_{-672}^{+896}$\\ 
102&103.817426&14.327622&2.2$\pm$0.6&213$\pm$7&3948$_{-617}^{+844}$\\ 
\hline
\end{tabular}
\end{tabular}
\end{table*}
\newpage
\begin{table*}
\caption*{Table \ref{tab:res249} continued.}
\footnotesize
\begin{tabular}{cc} 
\renewcommand{\arraystretch}{1.3}
\begin{tabular}[h]{cccccr}  \hline
Star & $l$ & $b$  & P$\pm$$\sigma_{P}$ & $\theta$$\pm$$\sigma_{\theta}$ & $d$$\pm$$\sigma_{d}$\\  
Id  &   ($\degr$) &    ($\degr$)&  (\%)    & ($\degr$) & (pc)\\\hline
103&103.822409&14.433230&1.7$\pm$0.4&183$\pm$7&1160$_{-46}^{+50}$\\ 
104&103.822561&14.466048&1.2$\pm$0.3&170$\pm$6& 962$_{-18}^{+19}$\\ 
105&103.824957&13.579997&2.9$\pm$0.5&179$\pm$5&1155$_{-68}^{+76}$\\ 
106&103.833996&14.484050&0.6$\pm$0.1&174$\pm$6& 682$_{-10}^{+11}$\\ 
107&103.834634&14.218497&2.0$\pm$0.3&183$\pm$5& 918$_{-31}^{+33}$\\ 
108&103.837299&14.291510&1.7$\pm$0.4&223$\pm$7&2621$_{-280}^{+352}$\\ 
109&103.841372&14.504933&1.6$\pm$0.2&172$\pm$3&4491$_{-510}^{+643}$\\ 
110&103.842941&14.199902&2.4$\pm$0.6&188$\pm$7&1252$_{-96}^{+112}$\\ 
111&103.848589&14.261173&4.0$\pm$1.0&157$\pm$7& 973$_{-41}^{+44}$\\ 
112&103.852653&14.485000&1.2$\pm$0.2&165$\pm$5& 742$_{-45}^{+51}$\\ 
113&103.855020&13.592059&0.9$\pm$0.1&182$\pm$3& 635$_{-275}^{+1194}$\\ 
114&103.858440&14.428986&1.3$\pm$0.4&181$\pm$8& 999$_{-45}^{+50}$\\ 
115&103.867833&13.513033&0.9$\pm$0.2&182$\pm$4&1503$_{-29}^{+30}$\\ 
116&103.872988&14.462072&0.8$\pm$0.2&163$\pm$8&1016$_{-26}^{+27}$\\ 
117&103.873111&13.833073&2.0$\pm$0.6&182$\pm$8&1538$_{-107}^{+124}$\\ 
118&103.882462&13.779338&3.0$\pm$0.5&192$\pm$4&1312$_{-152}^{+195}$\\ 
119&103.884349&14.449414&1.4$\pm$0.3&160$\pm$6& 968$_{-31}^{+33}$\\ 
120&103.884904&14.440311&1.7$\pm$0.3&175$\pm$4&1983$_{-128}^{+147}$\\ 
121&103.888410&14.261025&0.6$\pm$0.1&203$\pm$2& 553$_{-6}^{+6}$\\ 
122&103.889634&13.873195&3.7$\pm$0.4&213$\pm$3&2350$_{-191}^{+227}$\\ 
123&103.890580&13.579406&1.7$\pm$0.4&184$\pm$6&1476$_{-29}^{+30}$\\ 
124&103.894501&14.451970&1.5$\pm$0.2&173$\pm$4& 445$_{-4}^{+4}$\\ 
125&103.895567&13.555001&0.7$\pm$0.3&199$\pm$8& 761$_{-8}^{+8}$\\ 
126&103.896591&14.258742&1.2$\pm$0.3&207$\pm$5&6048$_{-1151}^{+1554}$\\ 
127&103.898571&13.623524&2.7$\pm$0.7&201$\pm$7&1569$_{-138}^{+166}$\\ 
128&103.901924&14.271198&1.1$\pm$0.2&200$\pm$4&2243$_{-150}^{+172}$\\ 
129&103.907345&13.604975&2.5$\pm$0.8&154$\pm$8& 518$_{-12}^{+13}$\\ 
130&103.912882&13.847550&2.3$\pm$0.2&225$\pm$3&1370$_{-37}^{+39}$\\ 
131&103.917862&13.901485&2.0$\pm$0.1&211$\pm$1& 478$_{-5}^{+5}$\\ 
132&103.920679&13.754134&1.9$\pm$0.5&201$\pm$5&8655$_{-1428}^{+1832}$\\ 
133&103.924358&13.587156&1.8$\pm$0.2&198$\pm$4& 800$_{-13}^{+13}$\\ 
134&103.925239&13.830119&1.5$\pm$0.2&219$\pm$3& 522$_{-4}^{+4}$\\ 
135&103.937410&14.509685&0.7$\pm$0.2&159$\pm$6&1066$_{-19}^{+19}$\\ 
136&103.941902&13.715247&1.2$\pm$0.2&199$\pm$3&1917$_{-48}^{+50}$\\ 
137&103.942720&13.852638&2.5$\pm$0.3&222$\pm$3&2783$_{-140}^{+155}$\\ 
138&103.943406&13.772898&2.1$\pm$0.4&200$\pm$5&1312$_{-37}^{+39}$\\ 
139&103.946746&13.767981&1.2$\pm$0.2&216$\pm$3& 849$_{-11}^{+11}$\\ 
140&103.949191&13.887192&2.1$\pm$0.2&226$\pm$3&3843$_{-403}^{+501}$\\ 
141&103.955544&14.539678&1.5$\pm$0.1&178$\pm$3&3345$_{-207}^{+235}$\\ 
142&103.970348&14.264221&4.6$\pm$0.3&125$\pm$2&2782$_{-571}^{+870}$\\ 
143&103.970485&14.478970&1.0$\pm$0.3&184$\pm$7&2569$_{-184}^{+213}$\\ 
144&103.976066&13.879128&3.2$\pm$0.7&225$\pm$6&3133$_{-345}^{+435}$\\ 
145&103.983313&14.557376&1.1$\pm$0.1&189$\pm$3&2313$_{-92}^{+100}$\\ 
146&103.987328&14.504984&1.6$\pm$0.1&183$\pm$1&1593$_{-62}^{+68}$\\ 
147&103.994239&14.075042&1.5$\pm$0.2&205$\pm$3&1825$_{-546}^{+1025}$\\ 
148&103.994364&13.897979&4.5$\pm$1.0&228$\pm$6&1769$_{-121}^{+140}$\\ 
149&103.996604&14.520752&1.9$\pm$0.5&178$\pm$7& 584$_{-17}^{+18}$\\ 
150&104.004217&14.059293&3.9$\pm$0.4&199$\pm$3&1832$_{-143}^{+168}$\\ 
151&104.007753&13.895467&3.0$\pm$0.6&228$\pm$5& 393$_{-4}^{+4}$\\ 
152&104.008566&14.116192&4.4$\pm$0.2&193$\pm$2&1413$_{-52}^{+56}$\\ 
153&104.011561&13.846133&1.5$\pm$0.3&221$\pm$6&1639$_{-57}^{+61}$\\ 
\hline
\end{tabular}
\begin{tabular}{cccccr}  \hline
Star & $l$ & $b$  & P$\pm$$\sigma_{P}$ & $\theta$$\pm$$\sigma_{\theta}$ & $d$$\pm$$\sigma_{d}$\\ 
Id  &   ($\degr$) &    ($\degr$)&  (\%)    & ($\degr$) & (pc)\\\hline
154&104.012635&14.040984&2.3$\pm$0.5&203$\pm$6&1071$_{-57}^{+64}$\\ 
155&104.013781&13.855258&1.9$\pm$0.1&225$\pm$2&2763$_{-154}^{+173}$\\ 
156&104.017669&14.484564&1.4$\pm$0.2&172$\pm$4& 572$_{-6}^{+6}$\\ 
157&104.022255&14.231824&1.9$\pm$0.5&159$\pm$7&1303$_{-90}^{+104}$\\ 
158&104.040345&14.029867&3.3$\pm$0.2&210$\pm$2& 484$_{-5}^{+5}$\\ 
159&104.040420&13.856945&2.7$\pm$0.9&219$\pm$9&1404$_{-38}^{+40}$\\ 
160&104.041785&13.895636&4.7$\pm$0.5&227$\pm$3& 441$_{-6}^{+7}$\\ 
161&104.042081&13.844076&2.7$\pm$0.8&230$\pm$8& 978$_{-22}^{+23}$\\ 
162&104.043729&14.115608&4.3$\pm$0.3&204$\pm$2&1105$_{-258}^{+460}$\\ 
163&104.049324&13.862234&2.3$\pm$0.4&225$\pm$5& 734$_{-17}^{+17}$\\ 
164&104.052854&13.842180&2.8$\pm$0.2&223$\pm$2& 831$_{-28}^{+30}$\\ 
165&104.061791&14.012051&2.2$\pm$0.2&199$\pm$2& 782$_{-12}^{+12}$\\ 
166&104.061949&13.859067&4.2$\pm$0.3&216$\pm$2&4179$_{-374}^{+451}$\\ 
167&104.068255&14.095157&2.3$\pm$0.3&194$\pm$4&2883$_{-293}^{+363}$\\ 
168&104.077033&13.878035&2.1$\pm$0.6&221$\pm$8& 654$_{-12}^{+12}$\\ 
169&104.081708&13.949807&3.1$\pm$0.5&209$\pm$4&1483$_{-64}^{+70}$\\ 
170&104.085362&13.965820&3.0$\pm$0.7&204$\pm$6&1569$_{-55}^{+59}$\\ 
171&104.095605&14.007224&4.0$\pm$0.5&213$\pm$3& 936$_{-38}^{+42}$\\ 
172&104.097804&13.977882&2.8$\pm$0.3&212$\pm$3&4534$_{-521}^{+659}$\\ 
173&104.100318&13.922871&2.1$\pm$0.4&215$\pm$5&1410$_{-37}^{+39}$\\ 
174&104.102274&14.013854&3.5$\pm$0.4&200$\pm$3&1102$_{-36}^{+39}$\\ 
175&104.104955&13.879472&2.7$\pm$0.9&225$\pm$9& 509$_{-10}^{+10}$\\ 
176&104.105705&13.906279&3.7$\pm$0.4&228$\pm$3&1721$_{-95}^{+106}$\\ 
177&104.114463&13.968833&1.5$\pm$0.5&222$\pm$7&1341$_{-79}^{+89}$\\ 
178&104.117886&13.985771&2.7$\pm$0.4&211$\pm$4&7276$_{-1184}^{+1546}$\\ 
179&104.128116&14.231034&1.9$\pm$0.4&157$\pm$6& 727$_{-26}^{+28}$\\ 
180&104.133295&13.852779&1.8$\pm$0.3&60$\pm$4& 1025$_{-19}^{+20}$\\ 
181&104.133852&14.389952&2.1$\pm$0.3&177$\pm$4&2857$_{-215}^{+252}$\\ 
182&104.141667&13.932187&3.3$\pm$0.6&207$\pm$5&1080$_{-27}^{+29}$\\ 
183&104.146638&14.520769&1.1$\pm$0.3&203$\pm$8&3023$_{-281}^{+341}$\\ 
184&104.159392&14.566307&0.8$\pm$0.1&165$\pm$4& 524$_{-4}^{+4}$\\ 
185&104.159795&13.895103&1.5$\pm$0.3&205$\pm$5&1230$_{-27}^{+29}$\\ 
186&104.166900&14.377326&1.2$\pm$0.4&171$\pm$8& 619$_{-15}^{+16}$\\ 
187&104.174435&13.912937&1.4$\pm$0.3&208$\pm$6&6703$_{-1126}^{+1493}$\\ 
188&104.177145&14.519745&2.1$\pm$0.4&166$\pm$5&8969$_{-1521}^{+1920}$\\ 
189&104.180551&14.510612&0.8$\pm$0.3&185$\pm$8&1245$_{-27}^{+29}$\\ 
190&104.184877&14.463091&1.8$\pm$0.2&179$\pm$3& 836$_{-14}^{+14}$\\ 
191&104.189132&13.884124&2.4$\pm$0.7&228$\pm$8&1019$_{-56}^{+63}$\\ 
192&104.190082&13.877895&1.6$\pm$0.1&199$\pm$1&2603$_{-140}^{+156}$\\ 
193&104.190936&13.956342&1.4$\pm$0.2&218$\pm$4&1899$_{-59}^{+62}$\\ 
194&104.196741&14.515530&1.5$\pm$0.5&183$\pm$7&3476$_{-433}^{+561}$\\ 
195&104.201883&14.420479&2.0$\pm$0.2&172$\pm$3&2426$_{-149}^{+169}$\\ 
196&104.203426&14.389243&0.8$\pm$0.1&180$\pm$4& 318$_{-2}^{+2}$\\ 
197&104.205921&13.888406&1.1$\pm$0.3&202$\pm$6& 818$_{-9}^{+10}$\\ 
198&104.206816&14.410361&2.3$\pm$0.7&166$\pm$8&4192$_{-664}^{+908}$\\ 
199&104.210431&14.400087&0.7$\pm$0.2&186$\pm$6& 306$_{-2}^{+2}$\\ 
200&104.217190&14.287866&1.7$\pm$0.5&52$\pm$7&  793$_{-11}^{+11}$\\ 
201&104.222302&13.960593&1.2$\pm$0.2&223$\pm$3& 957$_{-12}^{+12}$\\ 
202&104.229025&13.979337&1.0$\pm$0.3&208$\pm$7& 473$_{-4}^{+4}$\\ 
203&104.236310&13.952170&2.4$\pm$0.7&217$\pm$8&2034$_{-173}^{+208}$\\ 
204&104.237648&14.540158&1.2$\pm$0.2&163$\pm$5& 596$_{-7}^{+7}$\\ 
\hline
\end{tabular}
\end{tabular}
\end{table*}
\newpage
\begin{table*}
\caption*{Table \ref{tab:res249} continued.}
\footnotesize
\begin{tabular}{cc} 
\renewcommand{\arraystretch}{1.3}
\begin{tabular}[h]{cccccr}  \hline
Star & $l$ & $b$  & P$\pm$$\sigma_{P}$ & $\theta$$\pm$$\sigma_{\theta}$ & $d$$\pm$$\sigma_{d}$\\ 
Id  &   ($\degr$) &    ($\degr$)&  (\%)    & ($\degr$) & (pc)\\\hline
205&104.242881&14.500877&2.1$\pm$0.7&158$\pm$8& 393$_{-8}^{+9}$\\ 
206&104.259147&14.495868&1.3$\pm$0.3&168$\pm$6&1368$_{-33}^{+35}$\\ 
207&104.267690&13.917088&1.1$\pm$0.2&196$\pm$6&1366$_{-31}^{+32}$\\ 
208&104.270189&14.494670&1.4$\pm$0.2&180$\pm$4& 646$_{-7}^{+7}$\\ 
209&104.271671&14.537384&1.3$\pm$0.2&166$\pm$3&1937$_{-742}^{+1333}$\\ 
210&104.272463&14.479830&2.4$\pm$0.6&182$\pm$7& 462$_{-9}^{+9}$\\ 
211&104.274425&13.991687&2.3$\pm$0.5&215$\pm$6&2817$_{-188}^{+216}$\\ 
212&104.277089&14.354539&5.3$\pm$0.6&185$\pm$3&2245$_{-860}^{+1431}$\\ 
213&104.277266&14.450763&1.0$\pm$0.2&186$\pm$4& 758$_{-8}^{+8}$\\ 
214&104.278619&14.449882&1.1$\pm$0.1&185$\pm$2&3125$_{-172}^{+192}$\\ 
215&104.325274&14.085256&4.2$\pm$0.1&219$\pm$1&1902$_{-176}^{+214}$\\ 
216&104.325527&14.320067&3.8$\pm$0.6&190$\pm$5&7381$_{-1350}^{+1759}$\\ 
217&104.326000&14.281308&1.8$\pm$0.1&213$\pm$1& 725$_{-263}^{+788}$\\ 
218&104.327117&14.135960&2.4$\pm$0.6&210$\pm$6&3112$_{-244}^{+287}$\\ 
219&104.332801&14.304419&2.1$\pm$0.5&200$\pm$7&2125$_{-126}^{+143}$\\ 
220&104.334197&14.356322&2.7$\pm$0.2&205$\pm$2&1205$_{-25}^{+27}$\\ 
221&104.337399&14.017642&1.8$\pm$0.2&212$\pm$3&1622$_{-34}^{+35}$\\ 
222&104.342139&14.143725&4.3$\pm$0.4&203$\pm$3&1341$_{-77}^{+87}$\\ 
223&104.347367&14.418693&2.6$\pm$0.7&178$\pm$7& 825$_{-35}^{+39}$\\ 
224&104.348908&14.424935&1.3$\pm$0.3&172$\pm$5&1049$_{-24}^{+25}$\\ 
225&104.350080&14.363166&1.7$\pm$0.1&205$\pm$1&1778$_{-79}^{+86}$\\ 
226&104.351013&14.165030&2.3$\pm$0.1&204$\pm$1&2633$_{-91}^{+98}$\\ 
227&104.355040&14.404310&2.8$\pm$1.0&188$\pm$9&5245$_{-708}^{+922}$\\ 
\hline
\end{tabular}
\begin{tabular}{cccccr}  \hline
Star & $l$ & $b$  & P$\pm$$\sigma_{P}$ & $\theta$$\pm$$\sigma_{\theta}$ & $d$$\pm$$\sigma_{d}$\\ 
Id  &   ($\degr$) &    ($\degr$)&  (\%)    & ($\degr$) & (pc)\\\hline
228&104.356073&14.301893&1.8$\pm$0.5&204$\pm$7&1307$_{-51}^{+55}$\\ 
229&104.358815&14.413296&2.8$\pm$0.5&178$\pm$5&1058$_{-41}^{+44}$\\ 
230&104.359251&13.983480&1.3$\pm$0.3&219$\pm$6& 991$_{-20}^{+21}$\\ 
231&104.362494&14.132594&2.3$\pm$0.6&227$\pm$6&1516$_{-105}^{+121}$\\ 
232&104.369866&14.251045&3.8$\pm$0.4&197$\pm$3&2546$_{-884}^{+1433}$\\ 
233&104.371279&14.155014&3.4$\pm$0.5&198$\pm$4& 949$_{-127}^{+171}$\\ 
234&104.371715&14.402114&2.0$\pm$0.5&199$\pm$6&1246$_{-30}^{+32}$\\ 
235&104.374637&14.199342&1.6$\pm$0.2&220$\pm$4&1093$_{-24}^{+25}$\\ 
236&104.374790&14.115005&2.1$\pm$0.1&218$\pm$2&1744$_{-92}^{+102}$\\ 
237&104.375014&14.189853&2.5$\pm$0.4&207$\pm$4& 705$_{-17}^{+18}$\\ 
238&104.377829&14.145469&3.9$\pm$0.4&218$\pm$3&1165$_{-56}^{+61}$\\ 
239&104.385528&14.097980&1.1$\pm$0.3&222$\pm$8&2378$_{-129}^{+144}$\\ 
240&104.390020&14.071041&1.1$\pm$0.1&213$\pm$3& 940$_{-12}^{+13}$\\ 
241&104.392257&14.247356&2.6$\pm$0.4&200$\pm$4&3063$_{-158}^{+175}$\\ 
242&104.395140&14.329504&3.4$\pm$0.6&187$\pm$4&1274$_{-77}^{+87}$\\ 
243&104.404037&14.248927&3.0$\pm$0.3&205$\pm$3&2251$_{-82}^{+88}$\\ 
244&104.405975&14.334530&2.7$\pm$0.6&206$\pm$6&1076$_{-46}^{+50}$\\ 
245&104.408042&14.326883&2.1$\pm$0.6&190$\pm$7&1306$_{-86}^{+98}$\\ 
246&104.414166&14.191229&1.3$\pm$0.2&204$\pm$4& 888$_{-12}^{+13}$\\ 
247&104.416347&14.309724&2.5$\pm$0.8&203$\pm$8&1184$_{-34}^{+36}$\\ 
248&104.424996&14.165882&2.0$\pm$0.4&188$\pm$6&2279$_{-202}^{+244}$\\ 
249&104.430929&14.299678&3.5$\pm$0.7&206$\pm$6&2141$_{-211}^{+260}$\\ 
\hline
\end{tabular}
\end{tabular}
\end{table*}
\end{appendix} 
\end{document}